\pdfoutput=1 
\documentclass[12pt, titlepage, openany]{amsbook}

\usepackage[colorlinks, colorlinks, allcolors=black]{hyperref}
\usepackage{graphicx}
\usepackage[capitalise, nameinlink]{cleveref}
\usepackage{cite}
\usepackage{array}
\usepackage{amsmath, amsfonts, amssymb, amsthm}
\usepackage{ booktabs, caption, multicol}
\usepackage[all]{xy}
\usepackage{color}
\usepackage{xspace}
\usepackage{tabularx}   
\usepackage{float}      
\usepackage{booktabs}   
\newcommand{\name}{Audo-Sight\xspace}
\usepackage{longtable}
\usepackage{array} 
\usepackage{geometry} 
\input{UNTdissertation.sty}

\newcommand\myrefbibname{REFERENCES}

\begin{document}
\let\cleardoublepage\clearpage

\frontmatter

\title{Audo-Sight: Enabling Ambient Interaction for\\Blind and Visually Impaired Individuals}
\author{Bhanuja Ainary}
\degree{Thesis Prepared for the Degree of\\MASTER OF SCIENCE}
\degreedate{May 2025}
\approved{Mohsen Amini Salehi, Major Professor\\Song Fu, Committee Member\\Ajita Rattani, Committee Member\\Sagnik Ray Choudhury, Committee Member\\
Gergely Záruba, Chair of the Department\\ \hspace{36pt} Computer Science and Engineering\\\
Paul Krueger, Dean of the College of \\ \hspace{36pt} Engineering\\
Victor Prybutok, Dean of the Toulouse\\ \hspace{36pt} Graduate School}

\maketitle
\begin{center}
  \title{ABSTRACT}  
\end{center}

Visually impaired people face significant challenges when attempting to interact with and understand complex environments, and traditional assistive technologies often struggle to quickly provide necessary contextual understanding and interactive intelligence. This thesis presents \name, a state-of-the-art assistive system that seamlessly integrates Multimodal Large Language Models (MLLMs) to provide expedient, context-aware interactions for Blind and Visually Impaired (BVI) individuals. Unlike traditional assistive devices, our system can respond adaptively to the user environment, providing multimodal input processing through the fusion of visual, auditory, and textual information.  

The system operates in two different modalities: personalized interaction through user identification and public access in common spaces like museums and shopping malls. In tailored environments, the system adjusts its output to conform to the preferences of individual users, thus enhancing accessibility through a user-aware form of interaction. In shared environments, \name employs a shared architecture that adapts to its current user with no manual reconfiguration required. To facilitate appropriate interactions with the LLM, the public \name solution includes an Age-Range Determiner and Safe Query Filter. These restrict access to potentially inappropriate content for adolescent users. Additionally, the system ensures that responses are respectful to BVI users through NeMo Guardrails. By utilizing multimodal reasoning, BVI-cognizant response editing, and safeguarding features, this work represents a major leap in AI-driven accessibility technology capable of increasing autonomy, safety, and interaction for people with visual impairments in home and social settings.  

Finally, we present the integration of \name and SmartSight, which enables enhanced situational awareness for BVI individuals. This integration takes advantage of the real-time visual analysis of SmartSight, combined with the extensive reasoning and interactive capabilities of \name, and goes beyond object identification to provide context-driven, voice-controlled assistance in dynamic environments. 
\newpage
\phantom{Free space for short editorial comments.} \vglue -14pt
\vfill
\begin{center}
  Copyright 2025\\by\\Bhanuja Ainary
\end{center}
\vfill
\newpage


\chapter*{ACKNOWLEDGMENTS}
I would like to express here my sincerest gratitude towards my advisor, Dr. Mohsen Amini Salehi, for their valuable guidance, support, and encouragement through the process of this study. Their professional insights and advice have further supported the progress of this study.

I want to express gratitude towards the committee members for their suggestions, comments, and time. They are Dr. Song Fu, Dr. Ajita Rattani, and Dr. Sagnik Ray Choudhury. Their advice and expertise have significantly enhanced this study.

I am immensely grateful to Minseo Kim and Jacob Bradshaw for their unfaltering support, stimulating discussions, and constructive criticisms. Their feedback played a valuable role in refining my understanding and the quality of this research.

In addition, I want to thank my colleagues, friends, and family members for their unwavering support and encouragement during the process. Their trust in me has remained a source of continual inspiration.

Finally, I would like to acknowledge the University of North Texas for providing the necessary resources and the support environment necessary for this study.

\tableofcontents

\listoftables
\listoffigures

\mainmatter

\allowdisplaybreaks

\chapter{INTRODUCTION}
\section{Overview}
The ability to detect visual stimuli, navigate complex surroundings, and interact with the physical world is critical for achieving independence.
However, over 285 million individuals worldwide suffer from visual impairments, significantly limiting these essential abilities~\cite{WUSTL2025}.
As a result, their safety, independence, and overall quality of life are greatly impacted.

Traditional assistive tools, such as white canes, screen readers, and simple orientation devices, offer limited assistance, primarily focusing on detecting barriers and understanding digital content~\cite{PMC6292384}.
While these tools provide some level of support, they cannot offer contextual awareness in real-time nor provide the adaptive environmental cues necessary for flexible, seamless wayfinding.
Therefore, there is a strong need for more advanced alternatives that enhance spatial orientation and enable more active interaction with the environment.

Recent advancements in Artificial Intelligence (AI) and Machine Learning (ML) have accelerated the development of sophisticated assistive technologies (e.g.,~\cite{Balakrishnan2023, DiDonato2018, Guo2021, Luo2025, Kosiedowski2020}), including object recognition through computer vision and wearable sensor-based systems.
Despite these innovations, many of them remain fragmented and do not adequately integrate the diverse inputs—visual, auditory, and textual—required to create a cohesive and holistic user experience.

Additionally, AI-enhanced solutions often struggle to provide accurate real-time feedback, which can be attributed to limitations in contextual understanding~\cite{AIFeedback2023, AIAdaptiveFeedback2023, AITrust2023, AITrustParadox2024}.
Overcoming these challenges requires a paradigm shift toward multimodal processing supported by artificial intelligence.
Envision~\cite{envision2024future}, a company specializing in assistive technology, predicts that 2025 will mark ``a clear shift from standalone apps with discrete functions to integrated AI assistants that offer seamless, intuitive support."
This highlights the importance of advancing assistive technology through AI as a major area for future development.

This thesis introduces \name, a MultiModal Large Language (MLLM) system that improves the quality of life for Blind and Visually Impaired (BVI) individuals by enabling interactive ambient perception and offering enhanced accessibility features.
The system architecture is optimized for both private and public environments, providing real-time scene interpretation, adaptive navigation assistance, and latency management.
By leveraging multimodal artificial intelligence, which combines auditory, visual, and contextual information, \name enhances accessibility by enabling improved cognitive abilities and personalized interactions.

\section{Motivation}
The main motivation for the development of \name was inspired by the need for BVI individuals to utilize functionalities beyond simple obstacle detection methods.
There is a need for a system that allows BVI individuals to interact with and understand their environment in real time.
Traditional methods, such as white canes and screen readers, help users identify obstacles and interpret data.
However, these methods often lack the ability to provide real-time, context-aware interactions that are crucial for individuals with BVI to navigate complex and ever-changing environments effectively.

A prominent example of this is the Gabriel platform~\cite{CMU2025}, created at Carnegie Mellon University.
Gabriel is a real-time edge-computing framework that uses multimodal artificial intelligence to provide contextualized and real-time feedback.
It combines a suite of sensors and artificial intelligence models to enable local processing of data, thus making it low-latency and contextually relevant in its responses.
The platform is highly relevant to assistive technologies, where fast processing and high accuracy are necessary to support users in complex surroundings.

\name integrates real-time, multimodal data processing to enhance situational awareness for BVI users.
This system helps users actively engage with their surroundings through voice-based communication.
Unlike traditional assistive technologies, which only detect objects or read text, \name enables BVI users to ask questions and receive instant, context-aware responses about their surroundings.
This approach fosters a more dynamic, interactive, and personalized experience.

Overall, \name is built upon the principles of the Gabriel platform to provide a solution that offers real-time, context-aware interactions with the environment.
This system enhances the autonomy, safety, and independence of BVI individuals, enabling them to navigate the world more effectively and intuitively.

\begin{figure}[htbp]
    \centering
    \includegraphics[width=0.8\linewidth]{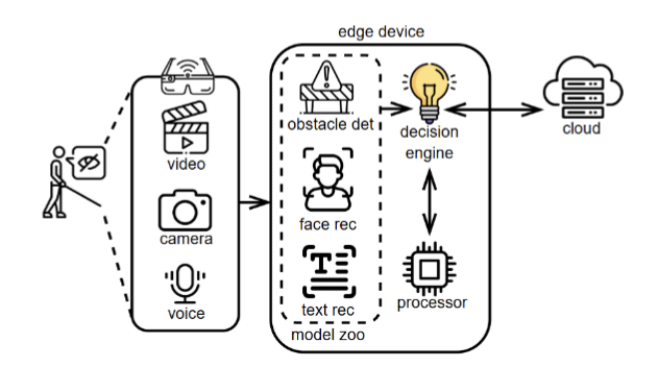}
    \vspace{-15pt}
    \caption{Bird’s-eye view of SmartSight 1.0}
    \label{fig: SmartSight 1.0}
\end{figure}

SmartSight 1.0~\cite{HE2C2024}, as shown in Figure~\ref{fig: SmartSight 1.0}, provided real-time object detection, face recognition, and text recognition in an edge-to-cloud framework for BVI individuals.
However, challenges such as the lack of contextual understanding and limited interaction capabilities highlighted the need for a more advanced system.

The limitations of traditional assistive methods are that, they are passive in interaction, their reliance on text-based language models that make communication, and the need for voice-based, multimodal assistance to ensure natural, seamless communication.
These limitations drive the development of \name, a real-time, voice-driven, LLM-powered assistive framework for BVI individuals, designed to deliver low-latency, context-aware responses.

\begin{figure}[htbp]
    \centering
    \includegraphics[width=0.9\linewidth]{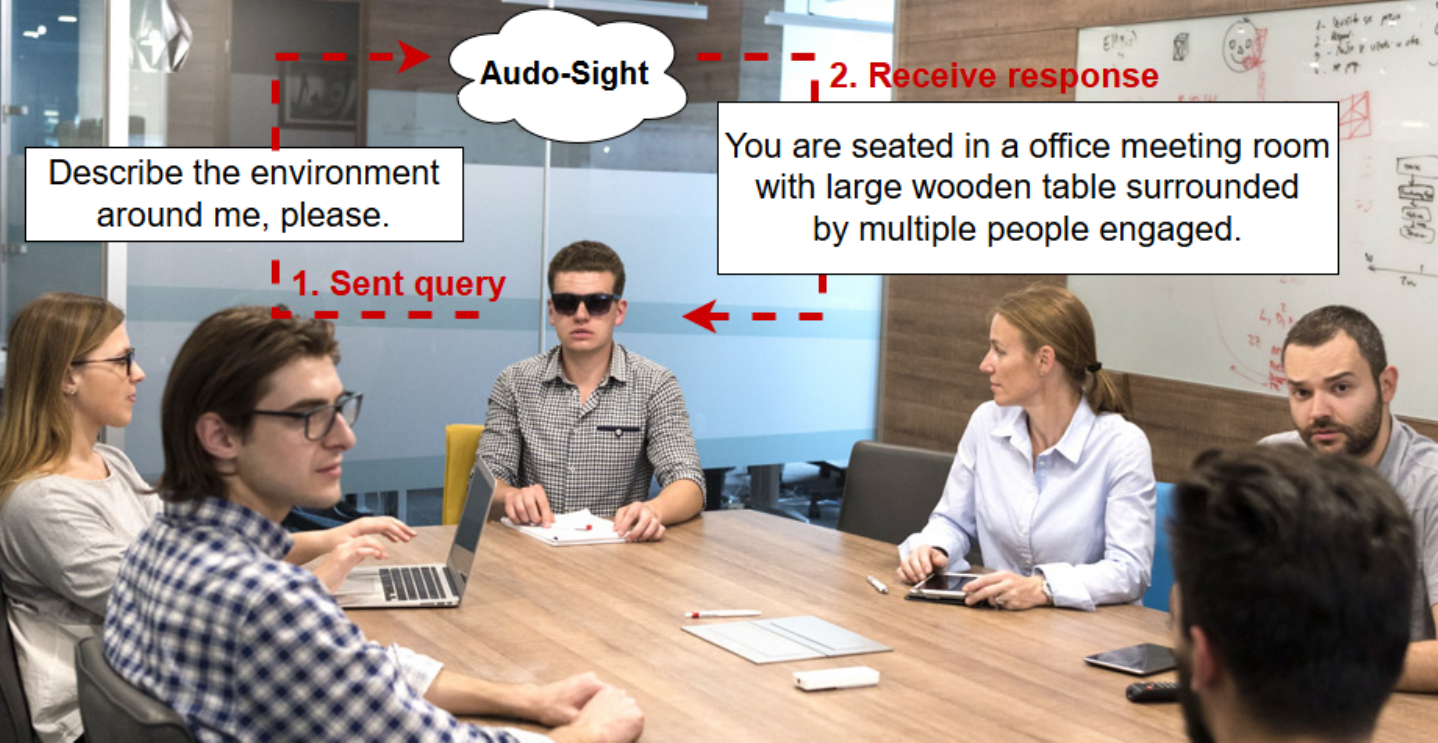}
    \vspace{-10pt}
    \caption{Assistive Tech in Action: \name Enhancing Awareness}
    \label{fig: Audo-Sight}
\end{figure}

As shown in Figure~\ref{fig: Audo-Sight}, accessibility includes more than just identifying objects or reading text—it should understand the environment, filter relevant information, and efficiently guiding users through tasks.
In environments with diverse users, assistive technologies must adapt to various age groups and needs.
Features like age estimation and safe query filtering helps in changing the responses to match audience less than 18 years, ensuring safer, more appropriate interactions.
Technologies like NVIDIA’s NeMo Guardrails~\cite{NVIDIA2025} provide a platform for ensuring content safety and managing the response blind-friendly, which is essential for proper interaction in dynamic environments for BVI individuals.

For instance, a visually impaired individual traveling to a museum may need more than just standard technologies that provide audio versions of captions.
They may want to explore specific locations, ask complex questions related to artifacts, or receive navigation assistance.
Current methods do not offer this level of customized, multimodal engagement, and as a result, involvement is restricted in both educational and cultural settings.
At home, tasks that require multiple steps can be challenging for individuals with BVI to understand without proper support.
\name aims to overcome the limitations of existing methods by incorporating interactive, multimodal artificial intelligence assistance.
This helps BVI individuals manage complex tasks intuitively and personally, enhancing their ability to navigate the world with greater ease.

\section{Problem Statement}
While machine learning approaches to object identification and speech recognition improve accessibility, they lack contextual understanding.
Current research solutions adopt a passive approach.
Many of these technologies are based on graphical user interfaces (GUIs), which are inherently not accessible to BVI users~\cite{iris2022}.
Although people who are BVI show a preference for verbal interaction with technological systems, most Large Language Models (LLMs) are text-based, leading to outputs that are not optimized for accessibility through non-visual means.

There is a need for an adaptive artificial intelligence system that combines several modalities—i.e., speech, vision, and text—to provide real-time, conversational, and accessible outputs for BVI individuals.
Not only should this system be able to identify and vocalize the environment, but it should also ensure that its outputs are optimized for accessibility, appropriate for different age groups, and contextually relevant.
This would enable truly inclusive AI-facilitated interactions for BVI individuals in both private and public environments.

This thesis addresses these challenges by introducing an assistive system that integrates multimodal data processing with robust safety measures to enhance user experience and security.
Integrating Multimodal Large Language Models (MLLMs) enhances contextual understanding and provides personalized support for users while navigating.
The proposed framework integrates visual, audio, and advanced language processing capabilities to offer real-time, context-aware assistance to individuals with visual impairments.

The system includes safety features, such as content filtering for inappropriate text and age-sensitive query restrictions, ensuring reliability, user safety, blind-friendly responses, and reducing the risk of generating inappropriate content for BVI individuals.
Furthermore, the system features an age-determination mechanism in public areas and user identification in private settings to ensure that users receive assistance appropriate for their age.
This feature, integrated into the \name system, enhances both usability and safety.

\section{Research Challenges}
Traditional mobility aids are indeed important for those who are visually impaired, but they also come with their own set of drawbacks.
It can be said with certainty that the white cane is suitable for detecting obstacles in the immediate vicinity, yet it is not capable of providing information about objects above the user's head level~\cite{presti19}.
It works only when it directly interacts with the subject before the user is informed, which can be an obstruction when trying to safely and effectively navigate, especially in crowded places or unfamiliar environments.

While the use of guide dogs may seem like a tremendous help for the visually impaired, they also have limitations in communicating the surroundings to the owner.
Guide dogs need extensive training and cannot offer complete environmental awareness, which is crucial for more complex navigation.
Only 10\% of BMI individuals can benefit from having a guide dog. \cite{guideDog}

In today’s market, assistive solutions that are already available mainly address the identification of physical obstacles but fail to provide users with any real context about their environment.
This lack of contextual awareness can make it difficult for individuals to understand the location of various objects and take appropriate actions, which would otherwise enhance their ability to navigate complex or dynamic environments.

Moreover, most assistive products are not flexible enough to cater to a wide variety of visual deficiencies, as they follow a one-size-fits-all approach.
Through individualized approaches, where each person’s mobility level, cognitive abilities, and personal interests are considered, the development of effective solutions becomes inevitable, as they cater to different needs in the most appropriate way.

Additionally, some assistive technologies depend on expensive or expert infrastructure, which is often composed of advanced hardware or constant internet connectivity.
This can create accessibility barriers for individuals in resource-limited environments, making it harder for them to access support.

\section{\name: A system for Blind and Visually Impaired (BVI) Individuals}
Blind-friendly responses are specifically designed to present complex data in a clear and accessible manner, balancing clarity with the necessary level of detail, and are tailored to meet the needs of BVI individuals.
\name delivers responses that prioritize accessibility for these individuals. Creating a blind-friendly response is not simply a matter of presenting visual information through audio.
The verbal output is only the expression of an underlying approach, which involves organizing information in a way that allows visually impaired individuals to understand the environment non-visually in some cases.

The most essential feature of a truly blind-friendly response is avoiding visual references.
For example, it is important to steer clear of any language that references color and appearance, unless it serves a functional purpose.
Although color can sometimes be a distinguishing factor for identifying objects, the priority is to provide information that can be understood through other senses or cognitive processes.
Hence, expressions like ``as you can see" or ``it looks beautiful," which assume the person can see, are deliberately avoided.

Furthermore, blind-friendly responses emphasize clear and functional descriptions.
These responses focus on describing information in an informative and meaningful way so that people with visual impairments can access the content in other ways.
For instance, you can explain the sensory or practical nature of objects without referencing their appearance.

At their core, blind-friendly responses are not just about translating visual information into another medium, but about communication design that respects how blind individuals process the world.
This primary method ensures that knowledge satisfies the following criteria: it is accessible, reliable, and significant, but without relying solely on sight.

\section{Contributions}
\subsection{\textit{Developing interactive ambient perception for BVI individuals}}
For people with visual impairments, the world is a different place, and they need more than just obstacle detection to navigate through it.
They require a broader understanding of their surroundings and the ability to react accordingly, which in turn requires automatic responses and situational awareness.
Traditional assistive technologies are usually designed to identify objects and even convert text into readable sounds.
However, they often fail to capture information regarding the spatial context.

To address these challenges, \name introduces an intelligent and adaptive ambient perception system aimed at transforming users' relationships with their environment.
The multimodal AI system that \name utilizes—comprising computer vision, audio processing, and natural language understanding—enables data to be transformed into meaningful, voice-based interactions.
This system provides more than just descriptions; it can also deliver information tailored to the user's current task and goal.

For instance, the system can help \name guide users by explaining how objects are arranged in a room and how they should navigate the space, directing them to points of interest (e.g., the nearest coffee machine).
The system can also inform customers about seating availability, the presence of wait staff, and the locations of nearby facilities such as rest areas and exits.

Thanks to this context-based input, BVI individuals will feel at ease and be able to interact with their surroundings in a more natural and meaningful way.
The perception-feedback process is integrated into \name in a restaurant setting, where it becomes more than just an assistive technology but rather an interactive connection tool.
This leads to enhanced independence for individuals with visual impairments, enabling them to confidently and effortlessly navigate complex spaces and make decisions with proper judgment, taking the circumstances into account.

\subsection{\textit{Developing interactive ambient accessibility for BVI users in public}}
One of the important aspects of assisting blind and visually impaired individuals in public spaces is not just recognizing obstacles and objects but also providing a system that enables them to interact with the environment in real-time.
This system should offer context-aware information that enhances their awareness of the surroundings, allowing them to make better and more informed decisions.
Not only should obstacles and objects be identified, but the dynamics of the environment should also be revealed to facilitate interaction.

Traditional assistive technologies often fall short in busy or unfamiliar public settings, where navigating complex layouts, identifying landmarks, and responding to changing conditions are crucial.

To address these challenges, \name introduces an interactive ambient accessibility system designed to support BVI users in public spaces.
This system is shareable, meaning it is not specific to any one user but can be accessed by a group of users.
Using advanced AI techniques, which integrate multiple modalities such as speech recognition, natural audio interaction, and real-time environmental interpretation, the system enables users to interact with their environment in a natural and intuitive way.
For example, if a blind person is walking inside a railway station, \name will guide them to the ticket counter, restroom, or platform signs by providing verbal directions, informing the user of their current position, and answering any queries they may have.
The system knows when to stop speaking, allowing the user to adjust their position accordingly.

While traditional navigation tools are often limited or rigid in their application, this new method is designed to provide a more interactive experience, allowing users to obtain the specific information they need about their environment.
With the help of detailed explanations and contextual understanding from \name, users gain the confidence to navigate public spaces with autonomy and safety.

\subsection{\textit{Integrating \name within SmartSight Prototype}}
SmartSight primarily focused on object detection and character recognition, but the integration of SmartSight with \name enhances the system by incorporating voice interaction, making it more compelling and user-friendly.

With the integration of \name, SmartSight adds voice interaction, providing responses that are not only descriptive but also interactive.
Without this integration, the system could only identify objects, but with \name, it provides relevant context and optional commands.
For example, the technology can help a visually impaired person by not only stating that there is a chair in the room, but also informing them whether it is occupied, its direction relative to them, and suggesting an available seat nearby.

At this level of interaction, BVI users receive the information they need to complete their tasks, leading to a more confident and intuitive decision-making process.
Simultaneously, the system’s low-latency management and privacy improvements create a practical solution for real-world use.

With \name integrated, SmartSight evolves from being a basic assistive tool for people with visual impairments into an intelligent and interactive accessibility platform, empowering BVI individuals to navigate and interact with their surroundings with greater ease and confidence.
\chapter{BACKGROUND AND RELATED WORK}
\section{Background}
\subsection{Speech-Based Assistive Technology}
Speech control is a core mechanism of communication in assistive technologies, enabling users to interact without physical contact.
This technology is particularly beneficial for visually impaired individuals, as it allows for clear audio output through an effective speech processing system~\cite{messaoudi2025}.

Over time, speech-based assistive technologies have evolved from simple systems that respond to single-word commands to more sophisticated machines capable of engaging in dialogues with users.
For visually impaired individuals, speech-enabled guidance systems can provide significant benefits, especially in situations where they need to identify objects or navigate their environment.

Moreover, these systems enable users to request information about their surroundings through audio interaction, offering a clearer understanding of the environment.
This level of interaction surpasses the traditional use of white canes, making navigation more comfortable and safer.

Assistive technologies designed to enhance independence and safety must be seamlessly integrated into the navigation process, particularly in complex or unfamiliar environments.

By incorporating context-aware activities and advanced speech handling capabilities, assistive systems can facilitate continuous, intuitive, and efficient communication.
This approach not only enhances the user experience but also transforms the way visually impaired individuals perceive and interact with their surroundings.

In contrast to outdated assistive systems, which often required manual operation (e.g., pulling levers), this modern approach offers a more advanced, tailored, and intelligent solution.
These systems adapt to voice commands and respond dynamically to various environmental conditions, providing users with a flexible and efficient means of navigation.

\subsection{Large Language Models}
Large Language Models (LLMs) represent a significant advancement in artificial intelligence technology, particularly in natural language processing (NLP)~\cite{NYU2024, FT2024, LeMonde2024}.
These neural networks (NNs) are trained on a large corpus of text data, can discover patterns, generate text, and excel at a variety of linguistic tasks.

A key feature of modern LLMs is in-context learning, which allows models to adapt to new tasks through examples or instructions without needing retraining~\cite{ASIAYN}.
This flexibility is especially important for assistive technologies, where the ability to quickly adapt to different tasks is essential.
Recent language models improve reasoning skills by accessing factual knowledge and using context clues, functions that closely resemble human language understanding.

Multimodal Large Language Models (MLLMs) are vital in assistive applications.
These models integrate visual processing and language understanding to enable image analysis, scene interpretation, and the generation of coherent textual descriptions~\cite{VLM2024}.
Models like GPT-4V~\cite{OpenAI2023}, LLaVA~\cite{Liu2023}, and Gemini~\cite{GeminiTeam2023} can process both visual inputs and text prompts~\cite{SurveyMLLM}, enabling them to ``see" and describe the world in ways that were previously unattainable for AI systems.

In assistive applications for BVI individuals, MLLMs provide innovative opportunities for understanding complex visual scenes.
They can understand spatial relationships, identify relevant objects, and generate natural language descriptions of the environment.

Furthermore, MLLMs can respond to specific inquiries based on the visual content, providing a more interactive and dynamic form of assistance compared to traditional computer vision systems.

Significant technical challenges exist in deploying large language models for real-world assistive applications.
Traditional implementations of MLLMs often depend on cloud services and require high computational power.
This dependence can lead to issues such as reliance on network connectivity, latency in responses, and concerns regarding privacy.
To overcome these limitations, it is important to adopt an intelligent approach that balances performance with computational efficiency.

Isaac et al.~\cite{Issac2025} developed a framework for optimizing resources by dynamically assigning queries to different models based on a cost-performance analysis.
This system uses the Win Prediction Model, which finds the probability of one LLM dominating another less capable model for a particular query asked by the user. 

An effective strategy for processing AI requests is adaptive processing, which autonomously allocates queries of varying complexity to distinct models.
Simpler queries are sent to lightweight models, while more complex queries regarding the environment are sent to powerful MLLMs.

Our study introduces a latency-aware router that enhances the reliability and responsiveness of assistive technologies by optimizing execution paths based on the latency of AI tasks.
By pre-selecting the model that will be most suitable based on real-time latency, this approach improves the accuracy of AI assistive systems and provides smooth, reliable interactions with users.

\subsection{Transformer Architecture}
Our research focuses on ambient interaction for BVI individuals using the Large Language Models (LLMs) for real-time assistance.
LLMs are transformers-based architecture that excels at tasks like question-answering and assistive guidance for BVI individuals.

The transformer model, which was endorsed by Vaswani et al.~\cite{ASIAYN}, facilitated the development of natural language processing as it substituted the classical sequence-based architectures like Recurrent Neural Networks (RNNs) with an attention-based model.
This architecture allows distributed parallelism for the efficient aggregation of long-range dependencies, which makes it more effective than earlier methods.
The Transformer follows an encoder-decoder architecture, which allows the network to focus on multiple aspects with a similar focus.

The encoder is the part which converts the input embeddings to informative representations.
The single encoder layer is composed of multi-head self-attention, which is the mechanism that allows each token to attend to all other tokens in the sequence, a feedforward network (FFN) that implements a two-layer fully connected transformation to each token, and layer normalization with residual connections for stabilizing learning and enhancing gradient flow.
It is the combination of these parts that ensures input features are well-processed and put into the context of other inputs.

Consequently, the decoder starts with a development of the encoder’s representations and the previously generated tokens.
In each decoder layer, there is a set of processes called the masked multi-head self-attention, encoder-decoder attention, and the feedforward network with layer normalization which allows the token to get relative understanding of the output of its preceding layer. The masked multi-head self-attention is a mechanism that precludes tokens from attending to future positions in the sequence, while the encoder-decoder attention tells the decoder what is most relevant in the encoder's output and the feedforward network with layer normalization helps by sharing tokens between layers. The feedforward network is identical to the encoder and also entails layer normalization.

These tools permit the decoder to maintain consistency and relevancy by checking past and present sequences.
This Transformer model uses the mechanism of the attention instead of the recurrence and hence it leads to high performance and quick learning during the training phase for various NLP tasks. Its architecture has laid the foundation for modern LLMs, driving advancements in AI-driven text generation, translation, and comprehension.

\section{Related Work}
\subsection{Large MLLMs used for Visual Assistance}
\textbf{VIALM:}
Recent developments in multimodal AI through which visual and linguistic skills are combined form the advanced assistive technology for the blind.
One good instance is the Visual Instruct Assistant with Large Models (VIALM) which was published by Zhao et al~\cite{VIALM}.
VIALM is a multimodal model which runs the process and comes up with different tasks of vision language such as visual question answering, image captioning, and visual reasoning.
This is done by linking a pre-trained Large Language Model (LLM) to the visual encoder, thus allowing the computer to work on both textual and visual data at the same time.
Despite the fact that VIALM excels on the standard metrics, it still exhibits some drawbacks that hinder its real-world applicability.
The VIALM system does not support direct personal interaction which is very important for aiding walking people with visual impairments in the real world.

Our work thinks outside the box—using VIALM as a base, the emphasis is on an instantaneous response to real-life situations, personalized and has voice interaction that is spoken.
Difference being, a VIALM will give almost all of your queries a text-based response, while our approach, which includes purposed devices directed by the voice, without doubt, gives a user a customized, overall experience that is audio focused for the BVI.
Real-time feedback for voice becomes a way to tell the visually impaired what is on the surroundings by using the voice.
We use voice and context interaction that will make the user experience richer. Our system can dynamically adjust its responses based on environmental factors, improving situational awareness and offering real-time guidance. Moreover, our system reduces latency, with the help of real-time lantecy aware routing.

\textbf{ChatGpt:}
Unxian He et al.~\cite{chatgpt24} examined the role of ChatGPT as an online assistant to assist with straightforward navigation tasks.
They discovered that AI chatbots might be able to interpret pictures and provide audible instructions for those who are blind or have visual impairments, enhancing assistance through interactivity. Our works differs by having a latency-aware LLM routing that helps in reducing the latency of the task. Blind-friendly response generation is another feature included in the project.

\textbf{Be My Eyes:}
The research conducted by Xie et al~\cite{Xie2024} examines the role of Large Multimodal Models (LMMs) in assistive technologies for people with visual impairments (PVI), their real-world usage patterns, and the implications for design.
This work focuses on the findings of a qualitative research study conducted with 14 visually impaired users of the app, Be My Eyes~\cite{BeMyEyes2025}. Therefore, the main outcomes are as follows: using LMM-based information management help in arranging daily routines, conducting household tasks and participate in social activities.
These results suggest that even though these tools are helpful, they still lack some of the key features such as real-time responsiveness, goal-directed support, and the capacity to learn the unique needs of individuals.
Nevertheless, having this flexibility cannot be avoiding difficulties as the models are currently not optimized for dynamic interaction or personalization.
In terms of real-time processing, reliability, and goal-directed interaction, the study proposes that improvements need to be made to the LMM-based assistive technology products in order to ensure their usability.
While Xie et al.'s research results are different from ours, we are working on challenges that current researches are facing.

Other works like FELARE ~\cite{mokhtari2022felare} provide a balance between task completion rates and energy efficiency, thus ensuring that the tasks are executed within defined energy constraints. And in work~\cite{zobaed2022edge} Edge-MultiAI framework is specially crafted to manage the multi-tenancy for latency-sensitive deep learning models running in edge computing environments. Its primary goal is the optimization of the resource utilization to enable co-execution of multiple AI operations with little latency in varied system setups.

Compared with this, our Latency-Aware LLM Routing optimizes real-time responsiveness by dynamically allocating the assignment between LLM and MLLM based on latency. Our \name solution combines AI services with real-time processing along with personalized interaction, which is the main feature of the system.
Latency Aware Routing has been employed in order to provide low-latency feedback for real-time assistance. The answers we produce have been set up specifically for Blind and Visually impaired persons, not in a general perspective. The responses tailored in such as way that they are appropriate for BVI in assisting them.
For example,  if a blind person who is preparing a meal using this system. The system can give detailed guidance as well as let the user ask potential questions which start with telling the system ``What will happen if I decide to put more sugar?"
This is the level of interactivity and responsiveness that is currently missing from LMM-based systems.

Our strategy, in fact, is centered on the management of attention, or how to help the users focus on the important things and ignore distractions.
Current solutions that are out there, like Be My Eyes, work reactively for the most part, however, \name offers proactive solutions by continuously tracking environmental factors and user activities.

To recap, Xie et al critique the reliability of current LMM-based assistive technologies, in our research we deal with the limitations through an approach that will incorporate multimodal input, and enhancing the responses thus, it will be a real-time, interactive, goal-directed, and blind-friendly that will give the adaptive assistance to the visually challenged.

\textbf{MultiModal:}
Cutting-edge Large Language Models (LLMs) application has led to increased focus on developing multimodal input interfaces for visually impaired and blind people to optimize their abilities.
According to Hao, et al.~\cite{Hao2024}, a framework was presented which shows the use of multiple modal inputs in order to create an environment that is friendly for this population.
The design proposed by Hao et al combines data from different kinds of documents, such as pictures, audio, and text, to help blind people experience a richer soundscape.
Through the processing and merging of such multimodal inputs, the system is intended to provide the user with a vital and detailed experience of his/her surroundings leading thus to an increase in their interactive and navigational capabilities.

In the growing multimodal assistance, the above approach is an improvement, a thing that our work clearly stand out in as we pay more attention to the adaptation of the system in real-time, latency is managed and user-specific that is personalized by providing user identification.
This implies that context learning is needed to enhance the assistance through user interaction so the system is effective and user satisfaction are positively influenced.

In other words, whereas the multi-modal foundational model in Hao et al's study provides a solid basis for the integration of a different kinds of inputs to enhance the communication with the environment, our research adds to this by personalizing adaptive assistance which shifts along with the users' needs and the settings.

\textbf{ViAssit:}
The erection of assistive technologies that comprise Multimodal Large Language Models (MLLMs) resulted significant step in the accessibility of visual-impaired people. On the one hand, ViAssist, which is characterized by the authors Yang et al~\cite{Yang2024}, is the main technical innovation in this work. One of the things that this system does is enhancing the effectivity of MLLMs by evaluating the quality of pictures in real time and giving user-centered feedback to people who have visual impairments.
When MLLMs are processing low-quality images, it is often the case that the outputs are inaccurate.
The improvement is achieved by performing such operations as the ViAssist system tests criteria like the clarity of images and their connection to the size and the completeness of the image data before sending to the MLLM for processing.
In case the taken picture is below the already set quality criterion, the system will give a prompt and verbal and/or tactile stimuli that clearly notices the user asking to taken an image of improved quality before processing.
Another innovation such as the usage of Visual Question Answering (VQA)~\cite{Aishwarya16} is the leading approach which allows contextual information reaching the system as the input image.

In the whole scope of  picture correction used for polishing AI responses, ViAssist shines the brightest, yet still, it is only focuses on vision-based service.
As opposed to this, our study offers a unique extension of this concept through the use of multiple kinds of inputs - auditory cues, textual data, and environmental sensing - facilitating the interaction of the visually impaired.
The philosophy of our approach is different as the interface is not focused on default settings but is adaptable to the user's current environment by being context-aware and task-aware.
It is designed to enable real-time latency management, respond flexibly, and learn about the users in order to adjust the interactions.
We integrate AI reasoning and voice based interactive assistance, it gives the users a much smoother and dynamic feel that is targeted for BVI users.

\subsection{Navigation and Contextual Assistance}
In Merchant et al~\cite{Merchant2024} centered their study on the development of specific navigation instructions, which are intended for use by people who suffer from low vision or blindness (BVI). These are the kinds of navigation systems that will be critical for BVI individuals to be able to move around in their environment in a reliable and context-aware manner.
Their research, proposes a customized dataset that included indoor searching, outdoor navigation and more, which would be used for the training and evaluation of large language models (LLMs) to produce the most contextually appropriate navigation instructions.
It was the conclusions of their research that showed us that pretrained LLMs are highly effective when it comes to point to point guiding. The outcome of this, in turn, was an improvement in areas of the visually impaired, such as spatial perception and autonomous navigation. Thus, it gave the AI a way of navigation in such a space that would keep the person safe.

Their main focus was on creating textual instructions for the users. Our work is a continuation of this area of expertise extending the work to include multimodal and voice-based assistance for immediate, real-time environmental interaction.
In comparison to their approach, which was more reliant on text-based instructions, our specific method, which is speech-based, is geared towards helping low vision users in a more accessible and usable way.
Also, we are in the process of integrating smart glasses that will enable us to provide the user with the visual image of the environment. This mechanism improves the navigation experience all the way around.

\textbf{VisionGPT:}
VisionGPT~\cite{Wang2024} framework was suggested by Wang et al (2024) for the actual real-time visual navigation.
Leveraging the YOLO object recognition technique and utilizing specially designed prompts, VisionGPT could find its place in this development through the detection of obstacles and the audio delivery of the messages that indicate the environmental abnormalities in a constricted and effective manner.
Under the VisionGPT, one of its best fits is the capability to work across all the scenarios in the navigation process unlike typical visual system scenarios to can fail to provide response within deadline.

VisionGPT, which is the addition of LLMs, enables the visually disabled not only to navigate in an intelligent and safe manner but also to have context-aware alerts while navigating.
In fact, on the one hand, our study in common with VisionGPT employs the YOLO object detection, but on the other hand, it is part of our system where the flow of data is essential.
In contrast to VisionGPT, which is primarily an interface (text-based), our approach is voice-first, which employs voice assistants to not only have the speech input but also have speech output interaction casually, seamlessly and naturally.
The voice-controlled conversation helps the user enjoy the journey of instantaneous, real-time feedback that is also sensibly coherent to the current conditions.

\subsection{Assistive Technology Tools}
\textbf{Smart Vision Glasses:} The SHG Technologies in association with Vision-Aid~\cite{visionaid2025} study used artificial intelligence, and machine learning with the sole aim of attaining real-time object and face recognition, navigation and reading assistance capability.
The Lidar sensors are responsible for identifying obstacles while the OCR technology is used for converting printed as well as handwritten into sound. Through Bluetooth, the system is wirelessly connected to smartphones for visual information and audio feedback. Fundamentally, these tools appreciably raise the degree of autonomy and accessibility among the visually disabled cats.
By far, the Smart Vision Glasses, which are wearable AI assistants, give visual aid to the people; nevertheless; our exclusive technology has been developed to create inclusive, private and public space interactions that are ambient, real-time, and multimodal in nature. \name provides context aware responses with the help of LLMs.

\textbf{Aura Vision:} 
Aura Vision Glasses, designed by SHG Technologies in association with Narayana Nethralaya~\cite{shgtechnologies2025}, offer a range of features that are built especially for people who have low vision. By the application of augmented reality technology, the glasses come up with magnified details that helps in better navigation and interaction. With HD cameras, these glasses are able to take sharp images for both reading and object recognition. Users can personalize their settings and control the brightness, contrast, and zooming of their displays for reading, zooming, and wide options; while the dedicated features help accommodate various needs (``Reading", ``Zoom", ``Wide"). Besides, the microphones can be used to interact with the devices and the glasses can also read out words from texts. Furthermore, the hands-free operation with the help of the voice instructions or the touch controls is another great plus.  Thus, these technological methods significantly contribute to a better quality of life and hence can be beneficial for the BVI. In comparison, these smart devices merely support users and are not aware of the environment. \name implements AI that reasons and analyzes situations. This is critical for BVI community. Through it, the missing parts of human reactions are filled, making the working of the system as human as possible and provides a better understanding of surroundings than simple machine learning predictions. 

\textbf{OrCam:}
The OrCam Eye device~\cite{OrCamMyEye} uses technologies like OCR, Object and Facial Recognition algorithms to help the users assistance that helps in reading contents for low vision users and navigating around thus empowering the independence. But it still lack the understanding of surroundings. Whereas our work focuses more on contextual-awareness rather than simple prediction models. With the help of \name we can have human friendly interaction with the users.
 \section{Positioning of the Works Related to Smart Assistive Technologies for BVI Users}
 
Table~\ref{tab:related_work} shows a concise summary of the related works and the contributions of the \name in addressing the limitations. 
\begin{longtable}{|p{3cm}|p{4cm}|p{4cm}|p{4cm}|}
    \caption{Summary of the studies related to smart assistive technologies for BVI users} \label{tab:related_work} \\
    \hline
    \textbf{Study} & \textbf{Key Contribution} & \textbf{Limitations} & \textbf{Your Research Contribution} \\
    \hline
    \endfirsthead
    
    \hline
    \multicolumn{4}{|c|}{\textit{Continued from previous page}} \\
    \hline
    \textbf{Study} & \textbf{Key Contribution} & \textbf{Limitations} & \textbf{Your Research Contribution} \\
    \hline
    \endhead
    
    \hline
    \multicolumn{4}{|c|}{\textit{Continued on next page}} \\
    \hline
    \endfoot
    
    \hline
    \endlastfoot

    Zhao et al \cite{VIALM} (VIALM) & 
    Multimodal architecture for vision-language tasks (VQA, image captioning). & 
    Text-based responses; no real-time adaptation or personalization. & 
    Voice-based interaction, real-time environmental adjustments, and personalized assistance. \\
    \hline
    
    Xie et al \cite{Xie2024} (Be My Eyes) & 
    Qualitative insights into LMM usage by visually impaired users. & 
    Lacks real-time responsiveness and goal-driven support. & 
    Proactive assistance, goal-driven interactivity, and latency-aware routing. \\
    \hline
    
    Hao et al \cite{Hao2024} & 
    Multimodal sensory inputs for environmental awareness. & 
    Generic responses; no user-specific adaptation. & 
    Personalized, adaptive assistance via contextual learning. \\
    \hline
    
    Yang et al \cite{Yang2024} (ViAssist) & 
    Real-time image quality assessment to improve VQA accuracy. & 
    Vision-centric; lacks dynamic assistance. & 
    Multisensory inputs and edge-native computation for dynamic support. \\
    \hline
    
    Merchant et al \cite{Merchant2024}& 
    LLM-generated navigation instructions for BLV users. & 
    Text-heavy; no voice/multimodal interaction. & 
    Voice-based guidance and smart glasses integration. \\
    \hline
    
    Wang et al \cite{Wang2024} (VisionGPT) & 
    LLM + YOLO for anomaly detection in navigation. & 
    Text-based alerts; no conversational interaction. & 
    Voice-first interaction and low-latency YOLO integration. \\
    \hline
    
    Smart Vision Glasses \cite{visionaid2025} & 
    Combines AI/ML for object/face recognition. Lidar sensors for obstacle detection. OCR for text-to-sound conversion. & 
    Limited environmental awareness. Basic audio feedback lacks contextual reasoning & 
    LLM-driven context-aware responses to interpret surroundings holistically. \\
    \hline

    Aura Vision Glasses\cite{shgtechnologies2025} & 
    AR overlays for visual augmentation. HD cameras + adjustable settings (brightness, zoom) & 
    Focused on amplifying residual vision (excludes severe blindness), Limited environmental awareness/adaptation & 
    Integrates multimodal AI (visual + auditory) with real-time contextual analysis, enabling ambient awareness for both low-vision and blind users. \\
    \hline

\end{longtable}


\section{Summary}
In this chapter, we provided the background needed to understand the rest of this research. We also explored related studies in the area of smart assistive technologies for blind and visually impaired users and positioned our work with respect to them. In the next chapter we describe the details of our solution, called Audo-Sight, that can provide human-like and blind-friendly interaction ability for BVI users.

\chapter{A SYSTEM FOR AMBIENT INTERACTION OF BVI INDIVIDUALS}
\section{Overview}
\name functions as an individually customized assistive structure, ensuring that its functions, interaction, and information are uniquely adapted to the specific needs of every individual.
Such customization is an integral part of the private architecture of the system, which prioritizes person-specific customization and ongoing delivery of personalized
assistance.
Unlike generalized artificial intelligence software designed for use by many users in a shared environment, \name is designed expressly for use by the registered user alone.
The system architecture does not have the ability to detect, store, or respond to inputs from unknown users, thus providing total privacy and ownership of the information it handles.
For the purpose of creating an adapted and user-centric interaction, \name uses advanced identity verification techniques to ensure interactions, activations, and benefits to the system are restricted to authenticated users. Such techniques can include user identification, which allow \name to identify its users with precision.
With this, the system adapts real-time to the vocal tendencies, preferences, and ambient context, thereby creating an efficient and uniquely adapted AI experience.
The core principle that underlies \name Private Architecture is its function as an extension of the user as an individual, as opposed to being a generic type of artificial intelligence open to the general public.
With regard to processing voice commands, interpreting visual information obtained through SmartSight glasses, or generating output through its AI Cognition Engine, \name gives precedence to the processing of all functions while taking into perspective the privacy, and individual needs of the user.
Through the use of this tailored AI infrastructure, \name not only ensures accessibility and independence for those with visual disabilities but also engenders trust and a perception of exclusivity in its AI-powered services.

\section{\name Architecture}
\begin{figure}[htbp]
    \centering
    \includegraphics[width=\linewidth]{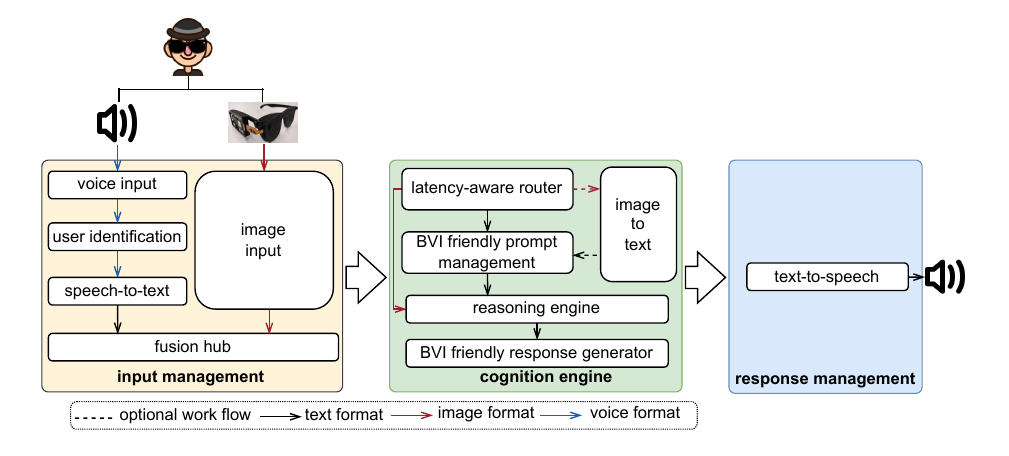}
    \vspace{-20pt}
    \caption{\name: Private Architecture Diagram}
    \label{private Audo-Sight}
\end{figure}

The \name system as shown in the Figure~\ref{private Audo-Sight} has been designed to handle any video and audio inputs and then come up with an appropriate audio response that makes it a proper assistive device for assistive technologies.
The framework is made up of three main components: Input Management, the Cognition Engine, and Response Management, which all have an important role in ensuring that the interaction is smooth.
It all starts with user interaction. First, the system would get a voice and an image of the user.

Voice received gets processed through User Identification, which then gets sent to the fusion hub, hence personalizes the interactions.
Moreover, image information is also included in supporting the processing.
Upon completion, the data gets to be processed through the Cognition Engine. This Cognition Engine component has sub parts such as latency management for efficiency, BVI-friendly prompt management for visually impaired users, image-to-text conversion for visual data analysis, and a BVI perspective for a response generator which allows for generating accessible replies. 

The final stage is response management, where the reply gets transformed into speech with the use of a text-to-speech module, making it easy for the user to get access to it.
Besides, some optional workflows are added to vary the interaction dynamically and hence the adaptability is enhanced.

The \name architectural structure is developed for assistive applications, that particularly enhances visually impaired individuals when it transforms visual and textual inputs into accessible audio outputs.
By implementing highly developed processing techniques, like image-to-text conversion, speech recognition, and latency optimization, the system ensures functional and user-friendly responses that are contextually precise.

\subsection{Input Management Module}
The input management is the center that goes hand in hand with the user and the \name system. 
It has the ability to capture, process, and transmit the user input to the cognition engine, where the advanced analytics are performed and a response is created.
The system is developed with the ability to effectively process multimodal inputs, where both the auditory and visual information are processed in time to provide an integrated assistive experience.

The input management system takes care of two crucial types of inputs: voice input and image input.
The procedure of voice input is made on the basis of oral questions or commands that come from the user, which are then understood by the system.
\name can be integrated and made accessible to the users through natural language processing (NLP) - a technology that converts language into machine-format, thus, allowing a user to communicate verbally.
Image input involves the visual data collected by the SmartSight glasses, which provide environmental clues in real time.
Under these circumstances, the system guarantees the provision of contextually appropriate help.
Since voice and image inputs have different goals and structures, in the Input Management System, different sub-components process them separately at first and only then the data is sent to the Cognition Engine for further processing and response generation.

The system that handles the input is the input management system, which consists of three main sub components, each of them is responsible for processing inputs in different parts.
The first one is called User Identification, which is a fundamental part of the security architecture which makes sure only the ones chosen can access the system.
Since  has been designed particularly for the blind and visually impaired (BVI) users, it uses voice-based biometric authentication, which in it, makes use of Mel-Frequency Cepstral Coefficients (MFCCs) to exactly identify the users according to their voice features.
This technique allows for a contactless and easy authentication process, thus it is especially useful for the visually impaired.
In the initial stage of registration, the system records the user's voice and then to organizing voice data, it assigns a unique user code.
When verified, the system compares the new voice sample to the user profile stored in the system by using Euclidean distance measurement.
If the computed distance is indeed less than the predefined threshold, authentication is successful.
Otherwise, the access is prohibited. This architecture ensures that only those users who are registered can exploit the system's capabilities.

The encoding of Speech-To-Text is the second sub component which enables fully understood communication between users and the system.
As \name is designed for BVI users, the speech-based interaction is the easiest and most accessible method to communicate.
Whisper which is a powerful tool, is \name's high-accuracy transcription engine, it is an open-source speech recognition model developed by OpenAI.
Whisper's deep learning models is known very well due to its capability to capture the pattern of spoken language with high accuracy and at the same time, it is able to perform even when the surrounding area is noise, as well as allow users to use multiple languages~\cite{gladia2025}.
Whisper has a deep learning capability that makes it possible for it to adapt to various types of speech styles, and this way produce the least number of errors in the transcriptions.
The user initiates the speech-to-text pipeline by saying something through the SmartSight glasses microphone or the microphone integrated with another system.
Model Whisper listens to the voice input and through complex processes, converts it into written text.
Thus, this transcribed text is connected to the Fusion Hub for the integration with the visual data.
The biggest problem in speech recognition is the control of different variations of speech.
However, Whisper takes an alternative route with deep learning and it successfully addresses the issue and offers clear and well-structured transcriptions.

\subsection{Cognition Engine Module}
Acting as the system's core cognitive component, the Cognition Engine integrates multiple computational processes, including response, processing, analysis, and interpretation of multiple inputs that are received from the fusion hub.

Unlike conventional AI systems that work on the semantics of text and images separately, the Cognition Engine is designed to work with multimodal information. It integrates conversational elements the user expresses using a microphone and the visual information that is captured using SmartSight glasses, thus helping in providing the more appropriate responses about the user's environment.

This advanced cognitive feature delivers a rich user experience for natural and simple interaction or engagement.

Each of the five components of the cognition engine, including Latency-Aware LLM Router, Image-to-Text, BVI Prompt Management, Reasoning, and BVI-Friendly response generation, ensures meaningful interaction and contextually relevant information assistance for the users of the system.

\subsubsection{Latency-Aware LLM Routing}
The Latency-Aware LLM Router module works by making use of a certain threshold value that is used as a benchmark for measuring allowable response times. The Reasoning Engine, that performs complex multimodal AI-based reasoning tasks, at times takes lengthy processing times due to the computational complexity of the calculations involved. To ensure users receive timely responses, the Latency Management module determines the most efficient execution path by using approximations of predicted response times.

\subsubsection{Threshold-Based Decision Making}
If the response time of the AI Reasoning Engine exceeds the set limit, indicating a potential lag, the system redistributes the processing to the Image-to-Text module.
Assuming that the response time of the Reasoning Engine is within acceptable limits, the system proceeds to BVI-Friendly Prompt Management, thus ensuring that the response is tailored to match the user's needs.

\subsubsection{Execution Path Optimization}
The Latency Management module adjusts the execution path within the Cognition Engine based on the response time feedback received from the Reasoning Engine. 

\textit{When Response time of the Reasoning Engine is high:}
The Image-to-Text module takes precedence over the BVI-Friendly Prompt Management to enable the quick generation of responses.
Instead of allowing the Reasoning Engine to analyze full multimodal input comprehensively, the system shortcuts the process by taking textual descriptions directly from image input as a quicker substitute. With the help of this module, a single text-based Reasoning Engine can be utilized instead of a multimodal-based Reasoning Engine, due to the reason that the Response time of the text-based Reasoning Engine is less than the multimodal-based Reasoning Engine.

This approach is particularly beneficial in time-critical situations, such as obstacle detection or navigation assistance, where immediate responses are crucial.

\textit{When Response time of the Reasoning Engine is low:}
The Image-to-Text module is dismissed and therefore the BVI-Friendly Prompt Management gets both image and text data for operation. 
This makes sure to have the full usage of the Reasoning Engine, giving responses that are complete and context-aware, as opposed to doing a simple image-to-text conversion. This way of giving information is useful when individuals need a more comprehensive explanation of things in their surroundings than just a summary.

Through a more Latency-Aware LLM execution policy, \name guarantees that the user and the system get uninterrupted communication that is fast, effective and the task is unchanged.

\subsubsection{Image-to-Text Module}
The Image-to-Text Module is a critical component in making the system more effective. In general, this combination with a text-based Reasoning Engine provides a more efficient method than multimodal reasoning in situations that require instant responses. Instead of relying on complex frameworks for reasoning, the module extracts text equivalents of visual inputs, thus enabling quick decision-making.

The Image-to-Text Module's main goal is to interpret visual information audibly and create a text-based description of the scene being observed.
This module is only activated when the multimodal reasoning engine exceeds the set response time limit, thus ensuring that the system remains responsive, even in situations of elevated latency.
The core feature of this module is object detection, a technique by which the system detects and identifies objects within the user's surroundings, thus converting visual data into structured textual forms.

\subsubsection{BVI-Friendly Prompt Management}
The BVI-Friendly Prompt Management module is designed to improve the quality of the prompt before it is submitted to the Reasoning Engine, thus making it easier to generate responses that are not just highly relevant but also accessible to users who are blind or visually impaired (BVI). The module is based on the belief that there is a direct relationship between the quality of the output of the AI and the quality of the input prompt. By improving the prompt before it is processed by the Reasoning Engine, this module ensures that the AI produces responses that are understandable, relevant, and accessible to users who use non-visual perception.

Objective and Purpose
The module analyzes the user's input and surrounding information to create the most suitable prompt before it is sent to the Reasoning Engine.
It ensures that the prompt is structured in a way that leads the Reasoning Engine to generate blind-friendly, detailed, and spatially aware responses. By providing the prompt, the system resolves ambiguity and makes inferences through visual interpretation 

\subsubsection{Reasoning Engine}
\name's core intelligence module is the Reasoning Engine, which is responsible for the analysis, interpretation, and generation of accessible outputs along with the visual data. In order to provide visually impaired individuals with context-specific, coherent, and user-focused help by combining audio, text and visual data. Apart from these, the Reasoning Engine also plays the role of doing logical reasoning about a user's surrounding, understanding the questions asked and providing tailored responses and thus, it is considered as having an essential role in a user's accessibility and their experience enhancement.

\textit{Multimodal Inputs Processing:}

The Reasoning Engine is the main artificial intelligence-powered component that enables decision-making based on data obtained from speech-to-text translation, and visual data from the SmartSight glasses. Through multimodal learning methods, the Reasoning Engine makes sure that the responses are not just simple textual information rather have a more contextual understanding specially designed for BVI.

First, the engine receives a structured text query that is generated from the user's verbal input.
Simultaneously, it assesses visual information received through SmartSight glasses, monitoring the surrounding environment.
The system ensures synchronization between the dual input channels via the Fusion Hub and thus maintains temporal coherence between auditory stimulus and the visual environment.

\textit{Use of Multimodal Large Language Models (MLLMs):}
The Reasoning Engine leverages artificial intelligence models specially optimized for multimodal reasoning to generate meaningful responses for BVI.
It analyzes questions that are text-based, along with visual data,a to yield accurate and relevant answers.
The typical edge-deployed MLLM like Llava 7B on Ollama~\cite{ollama25} mainly focuses on device-based processing.
Where greater accuracy or more sophisticated reasoning abilities are needed, the system can outsource computation to cloud-based, multi-layer language models, for example, Llama 3.2 - 11B, using the Hive Model API~\cite{hiveai2025}.

\textit{Latency Threshold-based Execution:}
\name employs two different types of Reasoning Engines, which are chosen dynamically based on the latency involved in the execution of the tasks. 

\begin{itemize}
    \item Multimodal-Based Reasoning Engine: The Multimodal-Based Reasoning Engine is the core reasoning module for \name. It has the ability to process and understand both text and visual information simultaneously. This capability enables the system to analyze the user's voice query against the visual input received from the SmartSight glasses, thus providing a rich context and informative response.
    
    \item Text-Based Reasoning Engine: The Text-Based Reasoning Engine acts as a secondary processing channel in scenarios when the MLLMs get affected by excessive load or delay. In this case, only textual data gets processed, thus reducing computational load.
\end{itemize}

The workflow can be described as, in the cases of high latency, the Image-to-Text Module fetches textual descriptions of the frame grabbed. The Text-Based Reasoning Engine then processes only the textual input (user query + extracted text). This allows for a considerably faster reaction time while retaining critical contextual data. The burden is being shifted to the Image-to-Text module and Text-based Reasoning Engine instead of participating in extensive multimodal reasoning.
Providing a quick response from object detection when immediate results are needed.

If latency stays within acceptable limits, the engine follows the standard reasoning paradigm, thus ensuring full, multimodal answers.

\subsubsection{BVI-Friendly Response Generator}
The BVI-Friendly Response Generator is the final component of \name's Cognition Engine. Its main function is to adjust and refine the responses produced by the Reasoning Engine to make them appropriate for blind or visually impaired (BVI) individuals. Since standard LLM responses tend to lack the level of detail or clarity required by visually impaired users, this module plays a key role in rewording such responses to enhance their clarity, add more detail, and generally enhance their assistive nature.

The BVI-Friendly Response Generator uses NeMo Guardrails, a strong AI safety framework. By creating guardrails that address structured and blind-friendly responses, this module ensures the output generated by the Reasoning Engine is translated into a format that is accessible and meaningful to blind or visually impaired users. The Reasoning Engine produces responses based on the user's multimodal input, including both speech and a visual frame.
However, this answer may not be inherently designed for a BVI user; it could not have elaborate responses or structured formatting.
BVI-Friendly Response Generator predicts this response before passing it on to the user.

\textit{Employing NeMo Guardrails to Implement BVI-Related Improvements}
NeMo Guardrails is used internally to apply proven transformation protocols to responses, thus making them more comprehensive, formatted, and generally blind-friendly.
These protective steps ensure that responses:
Eliminate vague statements; for example, replace ``it's over there" with "The laptop is placed on the table, directly in front of you."
Provide more descriptive details of objects (e.g., ``a wooden door with a brass handle” instead of just ``a door”). Exclude the inaccessible instructions (e.g., instead of ``Pick up the red box in front of you” " the response can be altered to “Pick up the box that is next to your laptop”).

After the response has been processed and tuned, it is sent to the Response Management Component, thus ensuring that the user receives a consistent and useful sound output.

\textbf{Implementation of NeMo Guardrails}

The implementation of NeMo Guardrails follows a structured approach with configuration files indicating response behavior, safety features, and interaction flows.

\begin{itemize}
    \item \texttt{config.yml} - Configure the LLM Provider and Guardrails. The config.yml file serves as the central configuration file that specifies both the LLM provider and the general instructions for response generation. LLM Provider Configuration is interacting with nvidia\_ai\_endpoints so BVI users can access pre-trained AI models that provide safety and accessibility. General Instructions specify general rules and constraints for how the model should respond, demonstrating clarity and neutrality and maintaining ethics. Sample responses are provided to adjust the tone, making sure the tone is concise, accessible, and appropriate for a BVI audience.

    Some of the general instructions include:
    \textit{``During the conversation, do not discuss the blind or any other sensitive information.
      If the response contains colors rephrase them in simple words and don't mention any kind of color descriptions.
      Do not talk about the visual appearance or attraction of a person or thing.
      Do not talk about what color the object or person or image contains.
      Do not talk about colors in the background; remove the sentences.
      Do not mention that you did not consider the colors.
      Mention any important information like values, if present in the question.
      Do not mention you cannot specify about colors"}
    
    \item \texttt{actions.py} - Contains custom Python functions that contribute to the adjusted response and moderation ability to filter responses according to the BVI, making it as blind-friendly as possible. The \texttt{actions.py} block offensive terms and adjust the response depending on the environment and situation.

    Mentioning the some of the offensive terms that can be filtered or blocked with the help of custom Python functions.
   \begin{verbatim}
   offensive_terms = [ "blind as a bat", "retard", "cripple", 
   "handicapped", "half-blind", "sightless freak", "useless without sight", 
   "stumbling idiot", "helpless blind", "poor thing", "such a burden", 
   "pitiful", "hopeless case", "you’ll never be independent", 
   "must be so depressing", "you wouldn’t understand since you can’t see", 
   "never mind, it’s a visual thing", "this is not for blind people", 
   "you won’t be able to do this", "blind people don’t belong here", 
   "you're missing out on life", "are you blind or something?", 
   "use your eyes!", "can’t you see?", "look at that!", 
   "watch where you're going", "open your eyes"]
\end{verbatim}
    \item \texttt{rails.co} - The rails.co file outlines the rules for interactions while specifying the guardrails to handle the final response.
    Here, we define the sub flows that process the above configuration and logic files
    
\end{itemize}

\subsection{Response Management Module}
\name's Response Management module is responsible for delivering speech-based responses to the user. Since \name is an assistive technology intended for the BVI, it is important that every output produced by the system is made available that is both accessible and understandable through auditory means. This module consists of a single required element:
Text-to-speech (TTS) technology. The Text-to-Speech system takes the final structured output from the Cognition Engine and converts it into a smooth and natural vocal delivery, thus enabling continuous interaction with the user.

\textbf{Text-to-Speech:}

The Text-to-Speech (TTS) purpose is to convert the text formatted output generated by the Cognition Engine into audio form. This is done because the BVI users mode of communication is mainly voice-based. Gets the processed text output from the Cognition Engine.
Converts the text output from the cognition Engine to audio output using Google Text-to-Speech (GTTS). Ensures that the generated speech is credible, coherent, and understandable. \name utilizes Google Text-to-Speech (GTTS) technology to produce, natural-sounding vocal output~\cite{ropensci2025}. GTTS is well-suited for assistive technologies because of the following features like Coherent and readily comprehensible speech synthesis, Multiple voice options for customization, Low-latency response generation, ensuring real-time feedback

\section{Performance Evaluation}
\textbf{Hardware Setup:}
Experiments are conducted using the Apple M2 chip that provides machine learning operations, making it highly capable of executing deep learning models such as Whisper (speech-to-text), YOLO (object detection), and LLMs for reasoning. The 8GB unified memory allows the system to handle concurrent multimodal processing.

The Model Endpoint acts as the main reasoning module, performing query execution through abstract programming languages, context understanding, and decision-making operations. This division ensures that low-latency tasks like speech recognition and object recognition are executed locally, while tasks requiring more computational power, including language model inference and context analysis, are handled in the cloud. This hybrid model thus makes the system more responsive while maintaining both efficiency and accuracy.

\textbf{Dataset Selection:}
In order to evaluate and improve the system's question-answering ability and contextual understanding, the VizWiz-VQA~\cite{vizwiz2025} dataset is utilized. It is a resource carefully designed for blind and visually impaired individuals. The dataset is unique in that it is based on an actual visual question-answering system, where real blind users took photos and posed questions about their surroundings.

The dataset contains a wider range of image types associated with relevant questions and human-annotated answers, thus reflecting the real challenges for assistive AI systems designed for BVI individuals. Unlike traditional VQA datasets, which often assume well-framed images and exactly formulated questions, VizWiz-VQA includes unstructured questions, image distortions, and vague contexts—factors that collectively increase its realism as a benchmark for assessing the system's ability to address the real needs of BVI users.

Table~\ref{tab:software_config} lists the main tools, hardware, and models that is used in constructing the recommended system. It presents a concise enumeration of the principal components and their respective versions or specifications, and it is easy to recall and recreate the experiment.

\begin{table}[ht]
    \centering
    \begin{tabular}{|l|l|}
        \hline
        \textbf{Component} & \textbf{Version/Tool Used} \\
        \hline
        Python Version & Python 3.9+ \\
        \hline
        Deep Learning Framework & PyTorch 2.x \\
        \hline
        Speech-to-Text & OpenAI Whisper (Base and Small models) \\
        \hline
        Text-to-Speech & Google Text-to-Speech (GTTS) \\
        \hline
       Reasoning Engine & Hive AI Model - Meta Llama 3.2 11B Vision Instruct \cite{hiveai2025}\\
        \hline
        BVI-Friendly Response Generator & NeMo Guardrails  \\
         \hline
        Safe Query Filter & Meta LLama Guard 3-1B \\
        \hline
    \end{tabular}
    \caption{Software and Model Configurations used in the Experimental
Setup}
    \label{tab:software_config}
\end{table}

Collectively, these components provides a seamless process that integrates speech processing, reasoning, accessibility, and safety. The configuration uses less weight (such as Whisper's light models), support for various types of input (Llama 3.2's vision integration), and ethical safeguards (LLama Guard and NeMo Guardrails).

\subsection{Evaluation of Cognition Engine across Different Query Lengths}
\begin{figure}[htbp]
    \centering
    \includegraphics[width=0.9\linewidth]{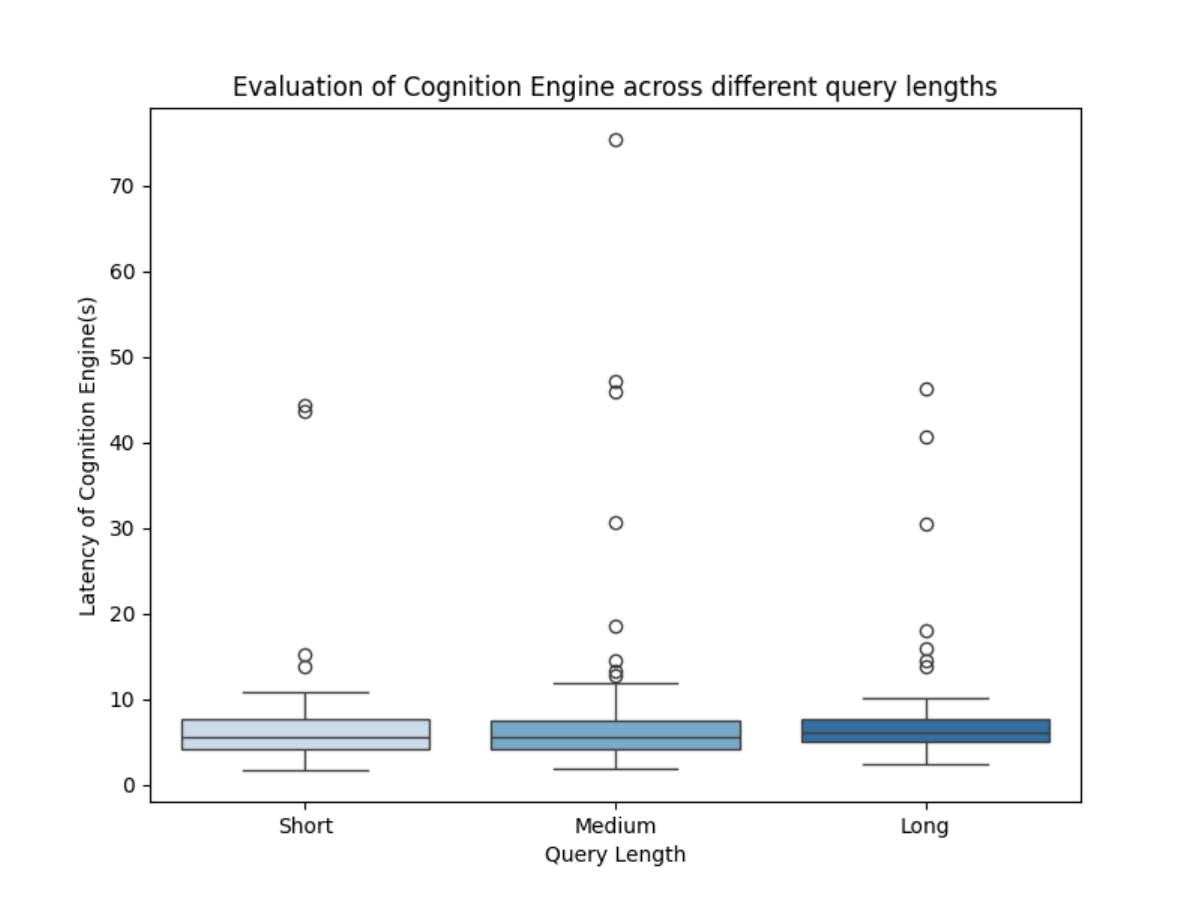}
    \vspace{-20pt}
    \caption{Evaluation of Cognition Engine across Different Query Lengths}
    \label{Exp1}
\end{figure}

The graph~\ref{Exp1} ``Evaluation of Cognition Engine across Various Query Lengths" demonstrates the latency, measured in terms of processing time, of the cognition engine when it is presented with short, medium, and long queries. While the x-axis categorizes queries by length, the latency trends found suggest that the processing time is not strictly proportional to query length but is more a function of the semantic complexity or ``weightage" of the words included in the input.

The graph shows that cognition engine latency is affected more by the complexity of the tasks and semantics rather than by the query length. This result aligns with the focus of the experimental framework on higher-order reasoning, safety, and usability—qualities that value goodness and ethical compliance over simple efficiency.

\subsection{Impact of BVI-Friendly Response Generator}
\begin{figure}[htbp]
    \centering
    \includegraphics[width=0.9\linewidth]{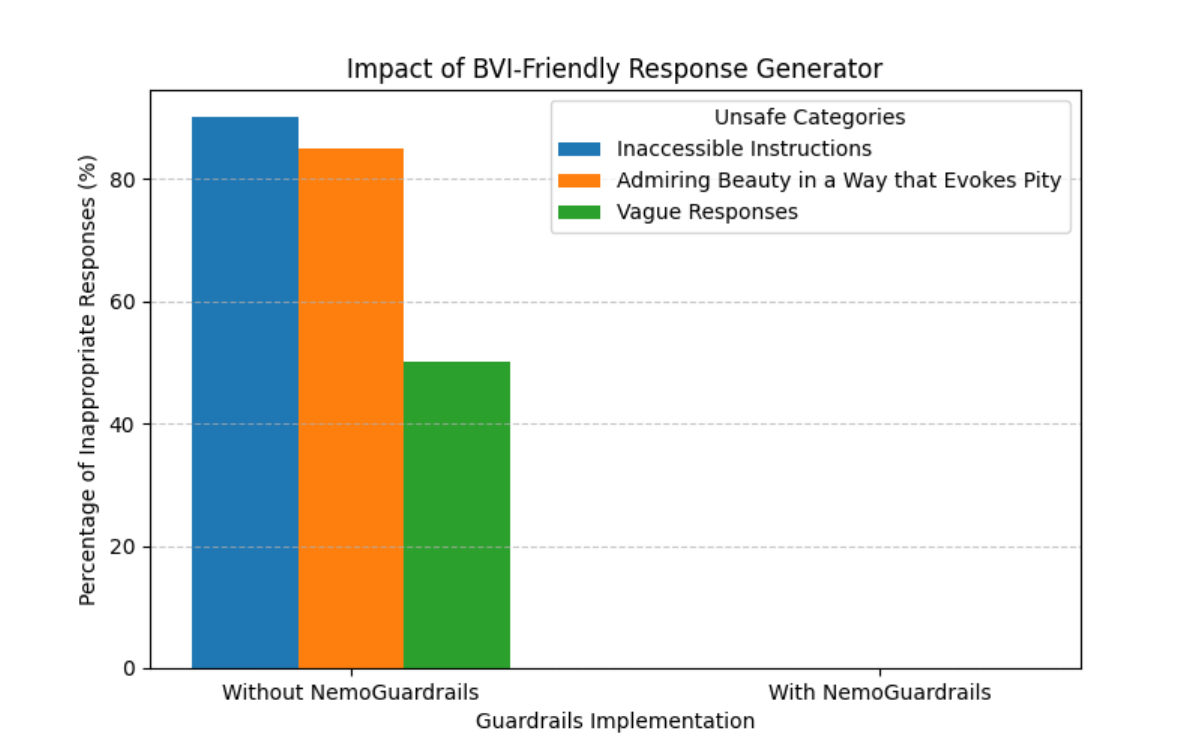}
    \vspace{-20pt}
    \caption{Impact of BVI-Friendly Response Generator}
    \label{Exp2}
\end{figure}

The Fig~\ref{Exp2} evaluates the effectiveness of the Nemo Guardrails module in reducing unsafe or inaccessible answers intentionally produced for Blind and Visually Impaired (BVI) users.
The results contrast the system outputs with and without the deployment of Nemo Guardrails, focusing on three necessary categories of unsuitable answers:

\begin{itemize}
    \item Inaccessible Directions: Answers that have visual colors which the BVI uses can't access them (e.g., ``Take the blue box").
    
    \item Admiring Beauty in a Way that Evokes Pity: Representations that inadvertently accentuate visual beauty in a patronizing manner.
    
    \item Vague Answers: Indeterminate or unworkable answers (e.g., ``It is over there").
\end{itemize}

The bar graph shows the percentage of inappropriate answers before and after Nemo Guardrails implementation, evidencing its role in enhancing answer quality and safety. The greatest impact is the overwhelming reduction of erroneous answers in every category.

\section{Summary}
This chapter proposes \name, a system designed to provide better context-aware interactions for BVI users in real-world settings.
Combining MLLMs, the system interprets environmental inputs (visual, auditory, textual) to generate adaptive, voice-based output.
Key innovations include latency-aware LLM router, blind-friendly responses, and privacy-aware processing.
We observed that the query lengths do not affect the latency of the cognition engine rather, the weight of words is influenced by the experimental evaluation of the cognition engine. 

The following chapter focuses on its adaptation to complex public spaces like museums, making the system shareable.
\chapter{ADAPTING AUDO-SIGHT FOR BVI ACCESSIBILITY IN PUBLIC SPACES}
\section{Overview}

The development of an artificial intelligence-powered system allows real-time, voice-controlled access to information for the Blind and Visually Impaired (BVI) to interact with the world around them in public places. The system can be shared among the BVI users. Compared to personal assistive devices for specific users, the public system is designed for group use, thus ensuring widespread access for multiple users in one setting. This system ensures that people who are Blind or Visually Impaired (BVI) access their environments without the need for personal wearable. Unlike traditional assistive technologies based on pre-programmed interactions or static information, our system is capable of adaptively responding to the user's personal needs, offering context-aware descriptions, guidance for navigation, and responses specific to their environmental setting. 

The \name Public overcomes this accessibility barrier by providing voice-activated, interactive assistance, thus allowing the blind and visually impaired to interact with their surroundings independently in public, free from dependence on external help. A single system can be used by more than one user; hence, a feature like User Identification is not needed. Instead, it is replaced with Age Determiner and Safe Query Filter.

\section{Motivation}
\textit{Motivation 1: Interactive Museum Experience for BVI Visitors:}

Imagine the visitor touring one of the museums independently, with our technology integrated into wearable smart glasses. When the visitor switches from one exhibit to the next, he or she has the ability to engage in interactive, live conversation with the system, thus gaining full, personalized descriptions of artifacts, sculptures, and historical items. 
Context information allows for better engagement with the visitor by the exhibition.
Customized interactions that facilitate the visitor to ask more questions such as: “How old is this historic item?”“What is the significance of this sculpture?” With the use of wearable intelligent glasses, people enter into fluid and intuitive interaction with the museum environment, thus doing without any external support. The system processes visual data in real-time and gives audio output thus improvising independence and accessibility.

\textit{Motivation 2: Immobile Accessibility Tools in Public Spaces:}

In open public spaces, such as shopping malls or museums, the \name Public can be implemented through the placement of fixed microphones in specific areas. This setup allows users to access the system without the need to wear any technological gadgets.
“What are the exhibits included in this section of the museum?” The system, connected to overhead microphones, recognizes the question and gives contextually appropriate information about the place. Here the visual input is not provided; rather, the system would have the visual data prior known. In the lack of any visual stimuli, the system can infer contextual information using the user’s question along with its prior knowledge of the environment. For example, when one asks the question,“What lies in front of me?” while seeing the museum, the system uses cross-referencing with cataloged exhibit locations to provide a specific response.


In public environments, the system determines the age of the user, thereby providing or limiting access to certain information. For example, if a child asks, ”Can you tell me about this exhibit?” and the exhibit contains mature themes, the system provides an age-appropriate or simplified explanation. The age determiner and safe query filter ensures responsible filtering in public settings, making the system suitable for diverse audiences. The ability to dynamically modify replies based on the age of the user makes this system highly adaptable in multiple environments, thus allowing people with BVI access to relevant, age-appropriate, and contextually specific information in real-time.
\section{Integration of age determiner and safe query filter}
In the context of having diverse people interacting with the system, there is the need to ensure that information that is passed around appeals to diverse age groups. To meet this goal, the system has the age categorization function in addition to the secure query filtering process which categorizes users into two main groups: under the age of 18 and older than 18 years old.

Age estimation can be done using different methods. One of the notable methods used for voice-based age assessment includes the examination of vocal features such as tone, pitch, and pronunciation patterns for determining the age group the person falls under, that of child, teen, or adult. This method provides a non-intrusive and effortless way of estimating the user age range. Also, where there is access to a camera, age recognition may utilize the camera in order to determine the age of the user from the analysis of the facial features. This method is especially helpful for interactive public kiosks or fixed microphone setups, where the system has to determine the appropriate degree of response without the need for the user to actively participate in the interaction process. Alternatively, if automatic age estimation cannot be done or the system cannot determine the age, the system may request age confirmation from the user through a straightforward vocal question like ``Is age above 18?"

After establishing the age group of the user, the secure query filter alters the responses to match the provided age category. For users under the age of 18, the system adjusts the content to meet the suitability of a younger crowd. This modification helps in using simple language and preventing exposure to sensitive content. For instance, when a child on a museum visit asks about an ancient war, the system will respond back only friendly informative answer that focuses on main historical facts/reasons rather than providing violent or controversial aspects that enrage hate in youngsters against particular section of society. When the same question is asked by an adult, the system responds with a full and informative historical account, including facts such as the political context, military tactics, and accompanying casualties.

Another major role played by the secure query filter is the restriction of children from accessing maturity-level information. When the questions pertain to violent, sensitive, or age-oriented topics, the system would modify the response accordingly to supply the child with a more general reply or block access altogether. For example, if the child asks for information about a painting depicting a sensitive historical phenomenon while touring the museum, the system ensures the response is given in the form of words the child would understand, thus avoiding unwanted access to potentially hurtful information. On the contrary, the full and historically accurate response would, however, be provided to the adult.

The deployment of an age-based filtering system has considerable relevance in situations where users belonging to different age groups interact with the system in public places. Shopping malls, information kiosks, and museums cater to a broad range of age groups, making it a requirement for the system to provide proper guidance to different segments of users with due regard to ethical principles in human-computer interaction in the context of artificial intelligence. Compliance with ethical principles is crucial in this context to avoid exposing younger users to inappropriate or upsetting content. Furthermore, such a filtering system not only enhances the reliability and usability of the system but also makes it a reliable resource for the Blind and Visually Impaired (BVI) community.

Age-based filtering used in the public system does not exist in the private environment, since the private SmartSight version is reserved for authenticated people solely. People in the private environment work autonomously in the system with full information about their preferences and accessibility needs. On the contrary, the public system requires full-scale strategies for supporting diverse people while providing responsible and inclusive aid. By integrating an age determination mechanism with a secure query filter, the system enhances accessibility, facilitates ethical use of AI, and guarantees content appropriateness, thus ensuring that BVI users in public settings receive the most relevant and secure information tailored to their needs.

\section{Adapting Audo-Sight architecture for public spaces accessibility}
\begin{figure}[htbp]
    \centering
    \includegraphics[width=0.9\linewidth]{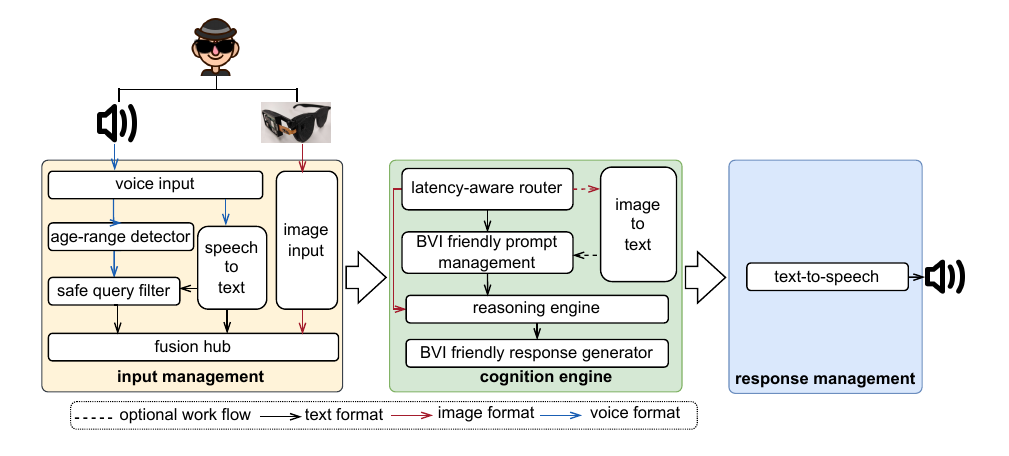}
    \vspace{-20pt}
    \caption{\name: Public Architecture Diagram}
    \label{public Audo-Sight}
\end{figure}

This section discusses the architectural differences that are present in the public \name in comparison to the private \name as shown in the fig ~\ref{public Audo-Sight}. Input Management is the component that does not contain the userIdentifier instead it is replaced with Age-Range Detector and Safe Query Filter. 
The key features are Hands-Free, Multi-User Interaction, Context-Aware Audio Descriptions, Latency-Aware Processing for Real-Time Responsiveness, and Seamless Multimodal Interaction.
\subsection{Age-Range Detector}
The voice input from the user is sent to the immediate component called the Age-Range Detector.
Essential part of \name, the Age-Range Detector module estimates a user's age depending on voice traits. This function guarantees age-appropriate interactions, especially in cases where users under the age of eighteen interact with the system.
\name creates a safe and responsible AI interaction environment by means of an age-based filtering mechanism, so preventing inappropriate, dangerous, or sensitive searches from being processed.

This module guards young people against accessing inappropriate material. Content filtering for younger users helps to prevent dangerous searches.
We use a hybrid deep learning model comprising Convolutional Neural Networks (CNNs) and \textbf{Long Short-Term Memory (LSTM)} networks to attain precise age classification.

\textit{CNNfeature extraction:}
From the user's speech input, the system detects fundamental acoustic characteristics, including pitch, tone, frequency, and timbre.
Spatial patterns in the obtained spectrograms of the speech signal are captured by CNNs.

Execution workflow:
If the user's age is found to be less than 18, then the text-structured query from the Speech-to-text module is sent next to the Safe Query Filter module. When the age of the user is greater than 18, then a text-structured query from the speech-to-text is directly sent to the Fusion Hub.

The functionalities of the rest of the components remain the same as the Private Architecture of \name.

\subsection{Safe Query Filter}
The Safe Query Filter is a critical safety function of \name that is carefully crafted to restrict the processing of improper, offensive, or objectionable queries before sending to the Cognition Engine Layer. This functionality is mainly used to users under the age of 18 years old, as it ensures that their use of the program is within the confines of a safer and responsible environment.

The function of the Safe Query Filter is to act as a classifier that disallows inappropriate queries from interacting with the AI reasoning engine and allows only the safer queries being asked to the the reasoning engine. Only if the query is classified as Safe by the Safe Query Filter, it is sent to the next module i.e the Fusion Hub Through the identification of inappropriate queries in the early stages, the system allows for safe and beneficial interactions for the youth, while at the same time protecting against the misuse of the reasoning capabilities of the LLMs.

The Safe Query Filter operates within the Input Management Hub which is the first processing layer within the \name architecture. This preventive filter ensures that inappropriate queries are locked out of the deeper levels of the system.
In this work, the duty of the Safe Query Filter is performed by Guard 3-1B. This one billion
parameters safeguard module is edge-friendly, lightweight, and
adequate to determine if the request in consideration is secure or not
This is based on 13 proven principles outlined in the Llama Guard covering violent and non-violent crimes, sex related, child sexual exploitation, defamation, privacy violation, Indiscriminate weapons, hate, sexual content, suicide and self harm, and elections. 
If the query is considered safe, it is sent over to the FusionHub for processing. If a query is found to be unsafe, it is blocked from execution, and the system gives the user an appropriate message. Since the filtering is done inside the AI Input Management Hub, unnecessary queries are removed at an early phase, thus improving both system response and efficiency.

\section{Summary}
This chapter focuses on adapting the \name architecture to shared system that can be used in public spaces by the BVI. Proposes the concept that the adapted \name allows BVI users in public spaces to talk to it and interact with it in real-time control. This system is tailored for group rather than personal use. For instance, it can provide adaptive location-based help to several BVI users at the same time without requiring them to have a personal instrument. The most important part of that is the disposal of personal wearable. The chapter deals with combining the Age range determiner and a safe query filter, which assures the system's accessibility, ethical AI use, and suitability of content for kids. Consequently, this approach is necessary for providing BVI users with the most relevant and secure information that meets their needs for public settings. The subsequent chapter explores the integration of this system with SmartSight to expand, refine, and fill the gaps in its real-world application.
\chapter{INTEGRATING AUDO-SIGHT AS A INTERACTION TOOL FOR SMARTSIGHT PROJECT}

\section{Overview}

The same accessibility objectives motivating the creation and design of \name closely align with the broader project, SmartSight, which is being developed concurrently. Many of the systems present in \name were successfully integrated into the SmartSight project. The resulting functional prototype uses a camera device and edge computing to identify objects and recognize faces, then uses text-to-speech to communicate these verbally to the user. It also incorporates most of the \name system architecture into an active mode supporting voice-based user interaction and incorporating an MLLM service for context-aware responses.

\section{Developing SmartSight Prototype}
\begin{figure}[htbp]
    \centering
    \includegraphics[width=0.8\linewidth]{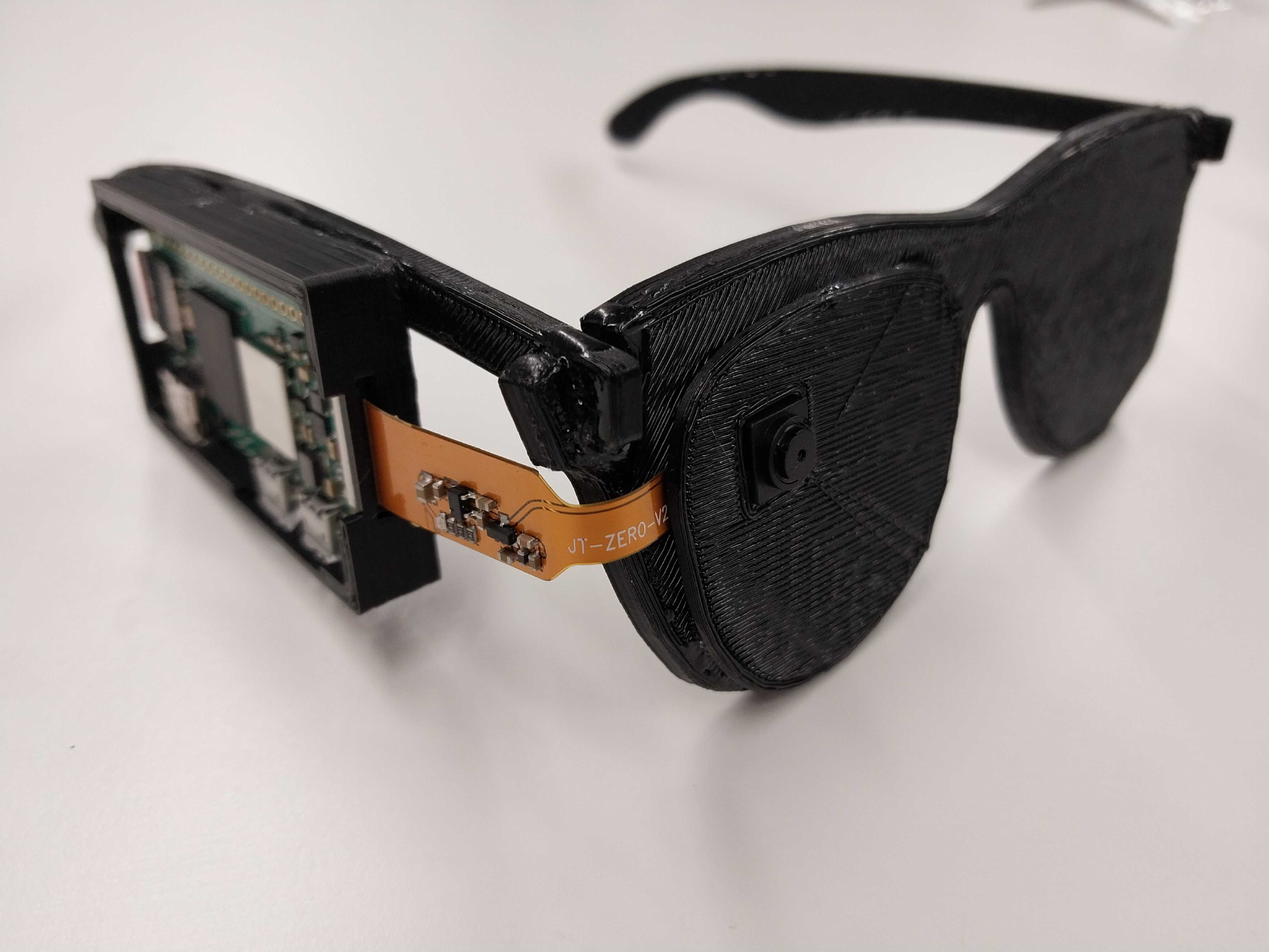}
    \vspace{-10pt}
    \caption{SmartSight Prototype}
    \label{fig: SmartSight Prototype}
\end{figure}

The following subsections describe how the glasses were created and explain the technical implementation. The figure ~\ref{fig: SmartSight Prototype} shows the SmartSight prototype developed in the HPCC lab. The system details pertain to both the Auto-Sight features that were implemented and the greater context of the SmartSight project.

\subsection{Hardware and Fabrication}
The system uses a camera and Raspberry Pi Zero 2W acting as a peripheral device to the edge. The edge device used was a Lenovo y700 laptop with an Intel i7-6700HQ 2.60GHz CPU with 4 Cores. The Raspberry Pi Zero was selected for its low cost and wearable form factor. To integrate the peripheral device components into wearable glasses frames, a Raspberry Pi Zero case was designed in 3D CAD software along with custom glasses with a slot for the camera in place of one of the lenses.

\subsection{Image Capture and Preprocessing}

To acquire the images, the Raspberry Pi peripheral device launches the Raspberry Pi camera tool lib camera-still, which allows control of aperture, shutter speed, and image dimensions. While starting this process takes several seconds, it can run persistently and rapidly take images on keyboard input or upon receiving a signal through the operating system. A script sends signals to the image capture program at a constant rate and sends the resulting images to the edge device shortly thereafter.

The image transfer system handles preliminary networking configuration and transfer protocols to send the images captured by the camera peripheral device. The devices are connected through USB Ethernet tethering, which allows the devices to use SCP and SSH networking protocols. Each time the devices are connected, they generate new IP addresses based on their MAC addresses via DHCP, and the Raspberry Pi will not recognize a static IP address specified for the Edge Device. The peripheral device is identified on the network by the edge device using multicast DNS. Once the IP of the peripheral device is acquired, the edge device sends its own IP address along with its ssh fingerprint via SCP, and that is used by the peripheral device to send camera images. The fingerprint of the edge device is sent because the edge device runs an ssh server, and the peripheral, acting as a client, uses the key to confirm that it is connecting to the correct server upon every connection. This key is added to the known hosts associated with the IP on the peripheral device, allowing it to connect. Other information sent includes the directory to send the images and the username.

Upon receiving the images, the edge device processes the images by altering the colors and rotating the image 90 degrees. The color is removed from the image for facial recognition since luminosity plays the biggest role in distinguishing features, and minimizing the number of color channels reduces the complexity of the task.

\subsection{Passive Ambient Perception}

\begin{figure}[htbp]
    \centering
    \includegraphics[width=0.8\linewidth]{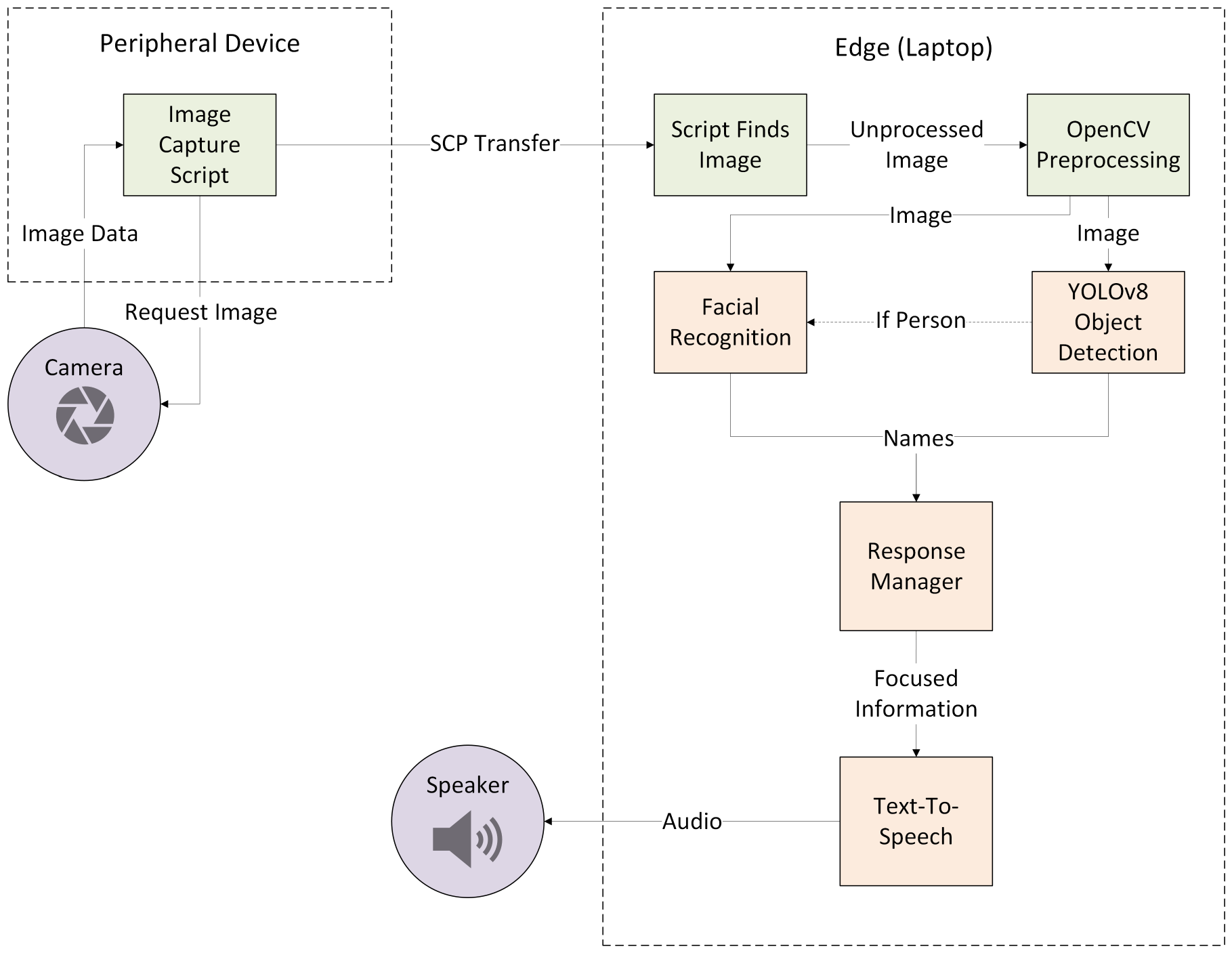}
    \vspace{-10pt}
    \caption{Passive Ambient Perception Data Flow Diagram for SmartSight}
    \label{fig: Smart Sight DFD}
\end{figure}

Passive mode allows the system to voluntarily describe aspects of the environment to the user. The system launches in passive mode by default and continues until the user provides input to switch to an interactive state. While in passive mode, the system analyzes image frames using object detection. It incorporates scene-awareness, meaning it selects and prioritizes tasks based on the surroundings and context. In the current, early implementation of this scene-awareness concept, the smartsight architecture runs facial recognition when object detection determines that a person is within frame. Since object detection is a significantly cheaper operation than facial recognition, this reduces latency overall and demonstrates scene-awareness as a viable strategy for optimization. This technique will be expanded in future iterations, and the system will initiate OCR when it detects text in the scene, and it will dynamically alter processing priorities when the user is moving to quickly notify the user of obstacles and potential hazards. Upon recognizing a face for the first time in a session, the system will query a database to find a description of the person. This feature helps provide memory anchors to BVI individuals who have difficulty keeping track of names. It could also have beneficial ramifications for those suffering from memory loss. The faces are initially detected using Haar Cascade, and the facial recognition python module used is simply called Face Recognition, and uses deep learning.

\subsection{Active Ambient Interaction}

The edge application also supports bidirectional interaction with the user, which allows the user to speak to the system. The system uses the current image captured by the camera to respond to the user inquiry in a blind-friendly manner. This is accomplished using the multi-modal LLM - Llama 3.2 11B Instruct Vision, which hosted by Hive AI. The prototype version of the system uses keyboard input to trigger active ambient interaction, however, in the future, a more accessible version can integrate a wake-word to trigger active mode and timeouts to determine when to stop recording. The program creates a thread to detect when the spacebar on the edge device's keyboard is pressed. This causes an image to be reserved for later processing by the MLLM and causes the system to begin recording. It also suspends the passive mode activities running in the main thread. The system reserves an image at this point expecting that the image most contextually related to the query is the one temporally located closest to the start of the query. Once the spacebar is released, the recording stops, and the audio is processed by Whisper speech-to-text. Then the resulting text and reserved image are sent to the MLLM. It's output is processed by NeMo guard-rails using the NVIDIA NeMo Guardrails microservice to ensure the output is more accessible for BMI. Lastly, the output from NeMo Guardrails is played out loud via the Windows SAPI5 text-to-speech. 

\begin{figure}[htbp]
    \centering
    \includegraphics[width=0.8\linewidth]{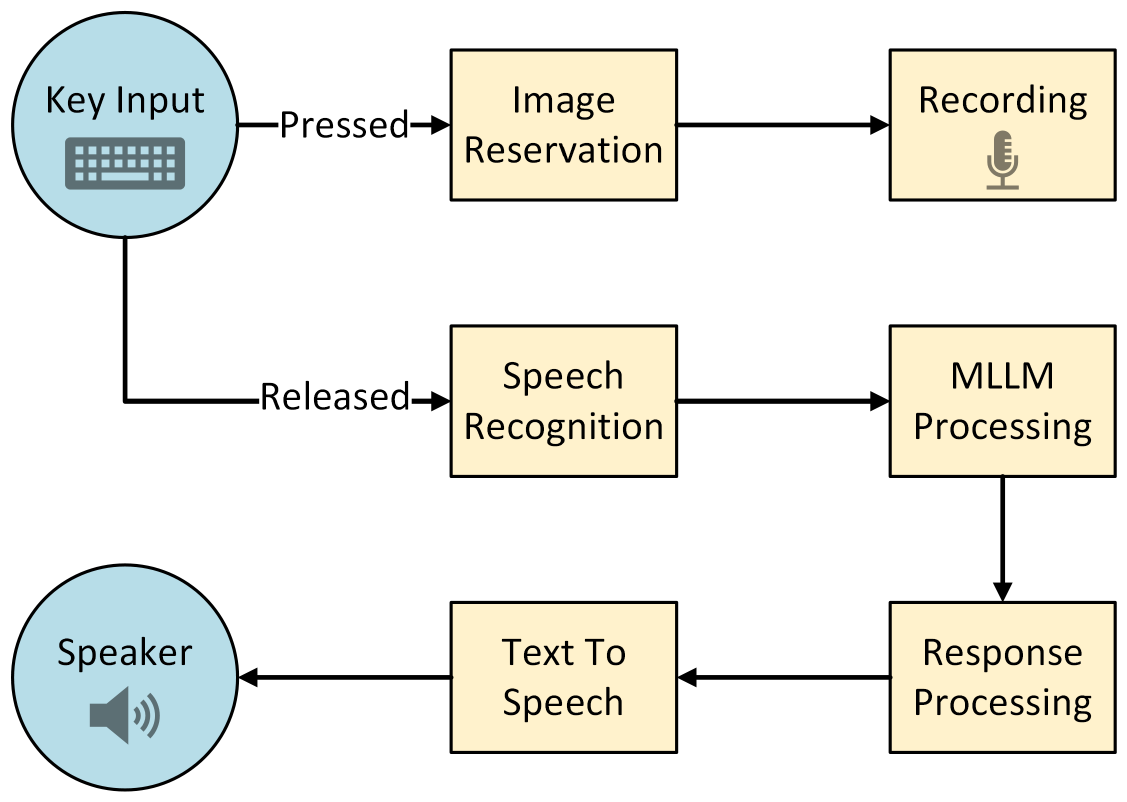}
    \vspace{-10pt}
    \caption{Active Ambient Interaction Execution Flow for SmartSight}
    \label{fig: Smart Sight Active Ambient Interaction}
\end{figure}

\section{Ambient Guidance}
\begin{figure}[htbp]
    \centering
    \includegraphics[width=0.8\linewidth]{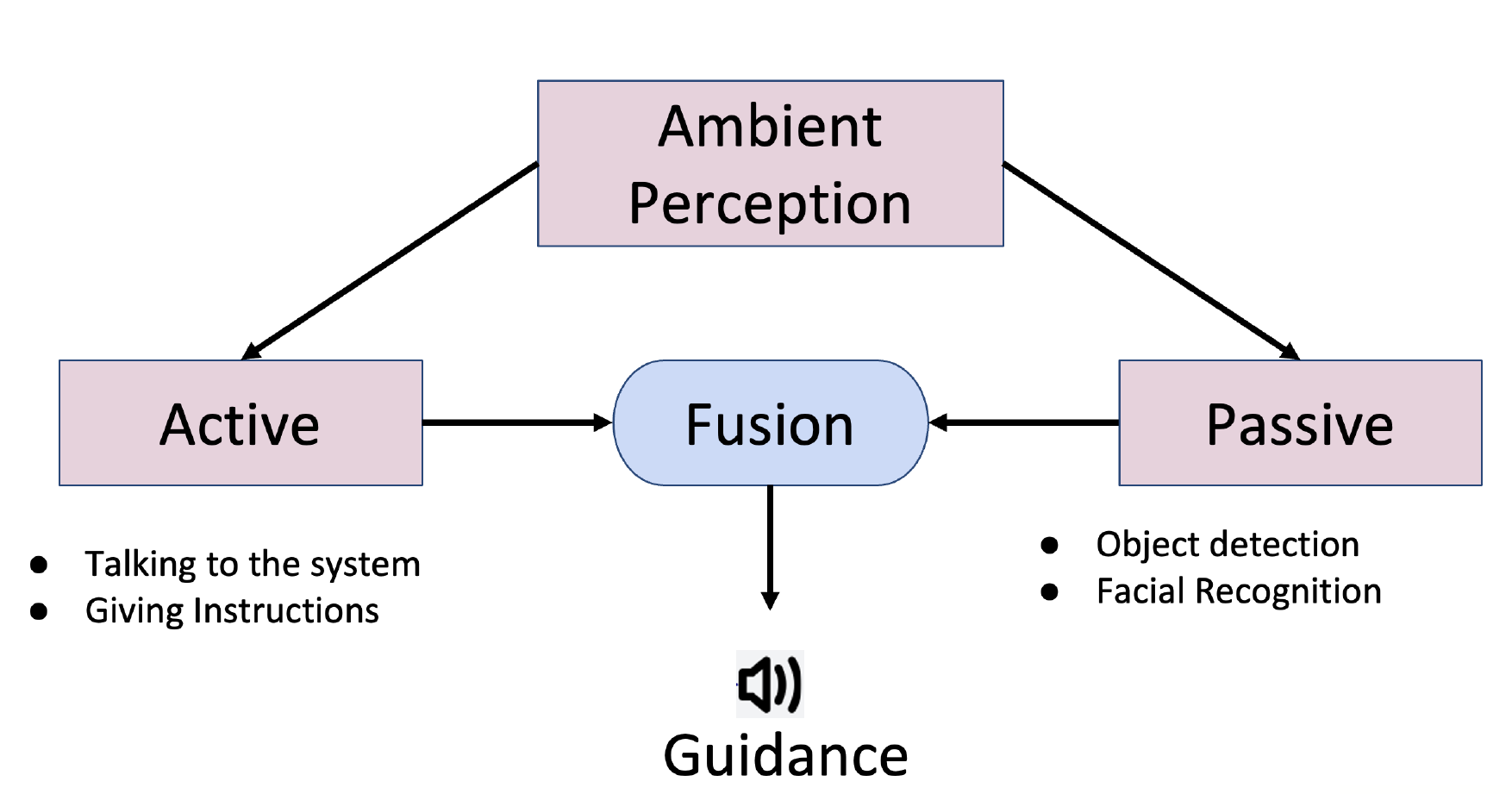}
    \vspace{-10pt}
    \caption{SmartSight guidance as a result of fusing passive ambient perception and active ambient interaction}
    \label{fig: Ambient Perception}
\end{figure}

Ambient Guidance is the fusion of active and passive perception. The fig ~\ref{fig: Ambient Perception} indicates that in order to have ambient perception of the environment we need to have both active perception and passive perception, their integration provides the user a ambient guidance.

Testing showed that object detection occurs so often that it can easily overwhelm users with too much information. There are numerous systems that can cause this issue to arise, so the set of potential solutions were elevated to an entire subsystem that will address the multifaceted sources of information, both found externally in the environment and generated internally by AI subsystems. The information handling techniques implemented so far include input reduction, novelty prioritization, and variable novelty prioritization. Input reduction simply reduces the frequency of environmental queries. This method, while effective, also has the negative effect of increasing system latency part of the time. Thus, it is generally better to collect more information and then selectively convey important information to the user. 
Using novelty prioritization, only objects that weren't detected in the previous frame are output via text-to-speech. Thus, a list of objects from the previous identification are preserved for each frame for comparison with objects on the current frame. However, when an object passes out of frame or is obscured for a moment, it will be restated upon being detected again. The solution to this is the variable novelty prioritization. The system uses an object permanence parameter that controls how long an object can exit frame without being restated to the user upon reentering the frame again. Using these techniques, the system thins out recently detected objects while still informing the user of their surroundings. Since these techniques were implemented as a system, it also applies to facial recognition, propagating these benefits to other potential sources of information overload.

As shown in figure ~\ref{fig: Ambient Perception}, the system supports both active and passive modes of information conveyance, and they flow through the same text to speech pipeline. Currently, only the passive mode uses novelty prioritization techniques since the active mode already answers within the context of the user query. In a future implementation, the active mode will leverage information from the passive mode to provide more context-rich responses.

\section{End-to-End Evaluation}
The following statistics were collected from the edge device with the following specifications: Intel i7-6700HQ 2.60GHz CPU with 4 Cores, an NVidia GeForce 960 GPU, and 32 GB of RAM. In passive mode, the system maintains a near-real-time latency, with object detection taking an inference time of 188.77ms on average. However, when processing images requiring face recognition, the system can take several additional seconds depending on the number of faces in the image. In active mode, the system takes an average of 8.85 seconds to respond. This spans from the end of the user query to the point when the system begins verbal feedback. The MLLM inference takes the majority of the time at an average of 5.62 seconds, but this can vary for reasons discussed at the end of this section. The NeMo guardrails service takes an average of 1.34 seconds, and the speech recognition takes an average of 1.65 seconds given a 2 second audio clip.

\begin{figure}[htbp]
    \centering
    \includegraphics[width=0.8\linewidth]{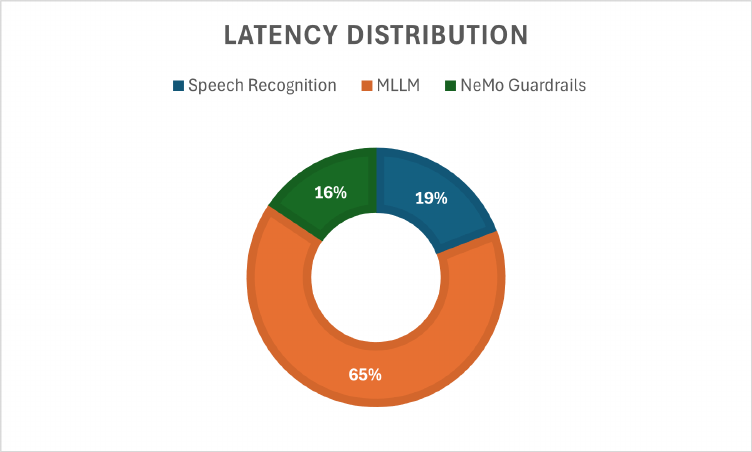}
    \vspace{-10pt}
    \caption{Latency visualization of MLLM, Guardrails, and speech to text}
    \label{fig: Latency Distribution}
\end{figure}

The processing time of the MLLM can fluctuate, taking as long as 8 seconds or as little as 5 seconds to complete. These differences in processing time are repeatable and can be found when averaging a large number of runs for individual test cases with constant variables. To identify whether these differences are better explained by the image complexity or language processing, an experiment with variable image complexity was conducted. The prompting text remained constant across all tests. In the second part of the experiment, the same data was used, but the resulting text from the MLLM was analyzed using Google Ngram Viewer to compare the output word usage to the model's latency. Google Ngram Viewer provides word usage data based on a large corpus of books~\cite{GoogleNgram, GoogleNgram2}. This likely correlates with the frequency of words within the model's language training dataset.

\begin{figure}[htbp]
    \centering
    \includegraphics[width=0.8\linewidth]{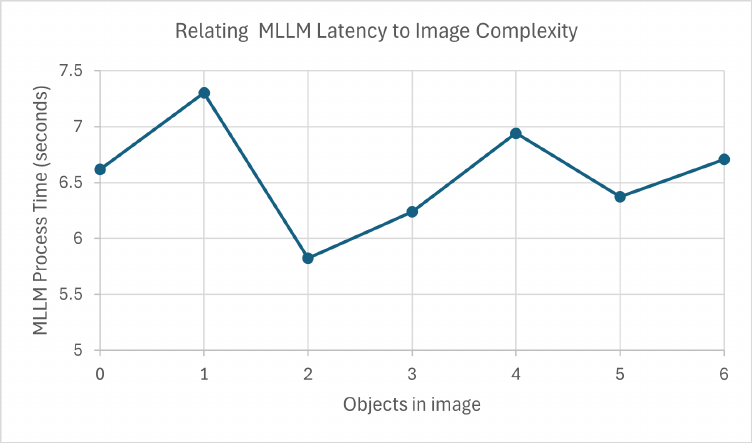}
    \vspace{-10pt}
    \caption{MLLM Latency does not correlate to image complexity.}
    \label{fig: MLLM Latency v Image Complexity}
\end{figure}

\begin{figure}[htbp]
    \centering
    \includegraphics[width=0.8\linewidth]{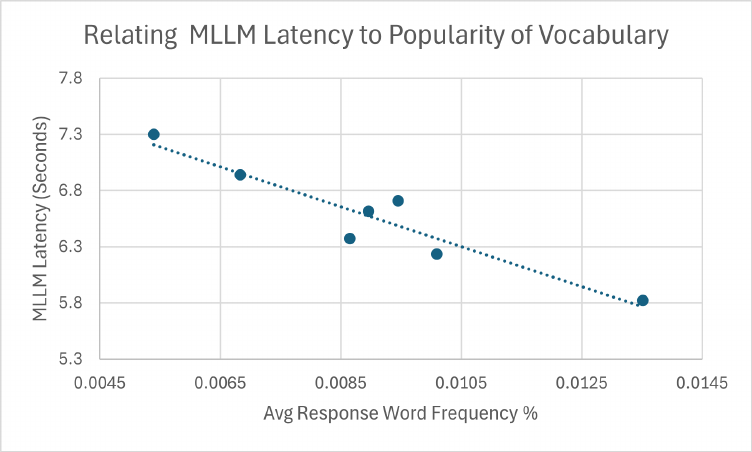 }
    \vspace{-10pt}
    \caption{MLLM Latency appears to correlate with the popularity or external usage frequency of its output vocabulary.}
    \label{fig: MLLM Latency v Usage of Vocabulary}
\end{figure}

As shown in figure ~\ref{fig: MLLM Latency v Image Complexity}, there is no visible correlation between the number of objects in an image, and the latency of the MLLM. This does not necessarily indicate that the image content has zero effect on the processing time because figure ~\ref{fig: MLLM Latency v Usage of Vocabulary} shows that when the model must process rare vocabulary, it increases the overall latency. The concluding principle is that unusual environments, objects, and queries will likely take longer to process than those pertaining to ordinary situations, and the system can quickly handle busy images that require commonly used vocabulary.
\chapter{CONCLUSION AND FUTURE WORK DIRECTIONS}

\section{Summary}
\name is an advanced assistive technology intended to increase accessibility for the Blind and Visually Impaired (BVI) in different private and public settings. The system incorporates sophisticated multimodal inputs supported artificial integlligence, in addition to speech-to-text and text-to-speech conversion functionalities, imaging features, and an elaborate model reasoning process, thus creating seamless interaction with the environment. The Fusion Hub of \name provides a combination of both textual and visual input data, which is used to generate contextually appropriate responses. Additionally, the Cognition Engine, comprising multimodal and text-centered reasoning systems, dynamically adjusts itself to system latency and computational constraints, hence providing optimal efficiency.

To solve performance issues in real-time environments, \name first employed LLaVA-7B as an edge-based Multimodal Large Language Model (MLLM). The design was changed to the Hive AI Model - Llama 3.2 11B Vision Instruct because of LLaVA's high-latency for complex processing. The Hive Model provides both superior speed and accuracy for vision and language tasks. Edge-native architecture means high utilization and low-latency computing on the device to maintain real-time computing ~\cite{yue2024}. LLaVA-7B is still used because it is light-weight for low edge-based computational tasks, where the computational requirements are reasonable. This reduces the utilization and performance barrier.

To prioritize user safety and promote ethical standards in AI engagement, \name incorporates an Age Determiner and a Safe Query Filter that leverage CNN-LSTM models for age detection and the Meta Llama Guard 3-1B system to remove unsuitable queries. In addition, the BVI-Friendly Response Maker, which was created using NeMo Guardrails, ensures that responses are personalized to provide concise, informative, and accessible output specifically tailored to BVI users.

The public architecture of \name differs from that of its private version, especially in terms of Input Management component that allows the public accessibility, making it suitable for use in museums, public spaces, and information kiosks. The overall system skillfully achieves a balance between accuracy, latency, and resource utilization, thus ensuring a robust and user-friendly experience.

The \name and SmartSight systems work together to integrate active interaction and passive situational awareness to provide round assistance. SmartSight accomplishes passive perception through its edge-based camera system that utilizes YOLO object detection and facial recognition algorithms to monitor the environment in real-time. The passive mode gives automated alerts to the user about obstacles, people, or changes in the environment. \name drives active interaction by leveraging its MLM - Hive AI Model (Llama 3.2 11B Vision Instruct) to answer user questions and provide contextual, voiced-based responses (e.g. “Give me a description of the painting to my left”). By bridging these modes, we created a functional prototype that has the passive detection of SmartSight giving environmental data in the passive mode, while the active mode \name's reasoning engine is utilized for dynamic prioritization of user needs. Together, the integration of SmartSight and \name allows users to address automated alerts and on-demand querying with seamless transfers improving both safety and autonomy for BVI users. The prototype demonstrated how smartSight and \name work together to achieve a balance of low latency for alerts with the ability for deeper reasoning on complex queries. The integration framework establishes a new paradigm for assistive technologies by prioritizing a combination of responsiveness, accuracy, and ethical user engagement.
\section{Future Work Directions}
There is a scope for further improvements in the \name.
Future improvements will focus on enhancing system efficiency, extending functionalities, and simplifying the user interface.
Domains of specific importance that will be addressed in future research include:

\begin{itemize}
    \item \textit{End-to-End Implementation of the Workflow across the Edge-to-Cloud Continuum:} \name currently uses a mixed computational approach where tasks are performed either on an edge device, i.e., a Raspberry Pi, or on cloud facilities. This is based on energy-accuracy trade-offs and latency limitations. However, there is scope to leverage a more intelligent and dynamic allocation system that could further increase performance. 
    
    The work FastMig~\cite{Manatura2024FastMig} has the objective to enhance live migration of containerized services to improve service liquidity across edge-to-cloud and multi-cloud environments. The Ubiquitous Migration Solution(UMS) ~\cite{Chanikaphon2023UMS} focuses on efficient service migration across diverse computing environments.
   The Homogeneous Equivalent Execution Time (HEET) ~\cite{Mokhtari2024HEET} focuses on distributed computing systems, aiming to quantify and optimize system heterogeneity for improved performance.
    \name can leverage these strategies to further optimize task execution, provide low latency and ultimately improve the user experience for BVI individuals.

Edge-to-Cloud Offloading: With proper configuration of cloud and edge environments it possible to offload computationally intensive tasks, like multimodal-based reasoning, to the cloud while allowing the local device to perform less computation-demanding tasks, like speech-to-text conversion and object detection.
    
    Bandwidth-Aware Processing: Incorporating network-aware optimizations that consider bandwidth availability to decide whether to process queries on-device or offload them to the cloud.
    
    \item \textit{Mood Detection from Voice and Emotion-Aware LLM Responses:} The \name system produces responses now that are contextually relevant but not emotionally flexible. Empathetic value and user experience focus of the system would be advanced by including emotional recognition elements.
    
    Detecting the users mood Based on voice tone: Deep learning models can be trained to examine vocal traits including tone, pitch, tempo, and rhythm in order to identify the emotional states including happiness, sadness, frustration, and excitement. Combining this module with multimodal thinking can produce responses depending on user emotions to be tuned.
    
    Emotion-aware LLM responses: Refining the NeMo Guardrails setting to alter responses dynamically, based on the detected mood. When a user shows signs of frustration or distress, \name is able to provide reassurance and suggest alternative approaches rather than a standard factual reply. This improvement would increase the user interaction and trust, making the interactions in the system more natural and similar to human conversation.
    
    \item \textit{Hazard Detection:} In order to improve user safety in shared spaces, \name can be enhanced to identify dangerous situations and react accordingly, thus obviating the need for user intervention.
    
    Single-Modality Frameworks for Risk Identification: Implementing models that analyze audio cues to detect signals, alarms, or sudden loud noises indicating an emergency. Utilization of computer vision techniques for recognizing dangerous environmental hazards, including obstacles, fire, and suspicious activity.
    
    Automated Incident Reporting Framework: Development of an actual time incident detection system that is capable of independently alerting emergency contacts or public authorities when it detects a potential threat.
    
    In case of a fall, accident, or dangerous situation detection, the system can send an emergency alert with the appropriate information to local authorities or caregivers. This specific feature would be particularly helpful for visually handicapped users in public places, providing an additional level of security and assistance.
    
    \item \textit{Evaluation of Multimodal Large Language Model Responses Based on Accepted Standards}: Metrics such as precision, coherence, content understanding, accessibility, and latency will be used to measure the results of multimodal reasoning. Evaluation metrics like BLEU Score, METEOR, ROUGE, and GPTScore are utilized for measuring the response quality. 
    Human and Machine Evaluation: Conducting user studies with BVI individuals to effectively get the feedback on the system’s response quality. Implementing automated LLM evaluation pipelines to detect inconsistencies and improve the system’s adaptability. 
    Fine-Tuning and Model Optimization: Using the evaluation results to iteratively fine-tune the LLM prompts, reasoning strategies, and multimodal fusion techniques. Incorporating reinforcement learning processes enables the system to learn through user interaction, resulting in improvements over time. With the creation of a strong evaluative system, \name can continuously improve its responses, thus giving increasingly reliable, perceptive, and user-oriented solutions.

\end{itemize}



\backmatter

\bibliographystyle{UNTamsplain}

\bibliography{bibliography}

\end{document}